\documentclass{article}

\usepackage{lineno,hyperref}
\usepackage{graphicx}
\usepackage{here}
\usepackage{amsmath,amssymb,amsfonts,bm}
\usepackage{booktabs}
\usepackage{url} 
\newcommand{\argmax}{\mathop{\rm arg~max}\limits}
\newcommand{\argmin}{\mathop{\rm arg~min}\limits}

\usepackage{enumerate}
\RequirePackage[OT1]{fontenc}
\RequirePackage[numbers]{natbib}
\RequirePackage{arydshln}
\RequirePackage[flushleft]{threeparttable}

\title{Spatiotemporal analysis of urban heatwaves using Tukey g-and-h random field models}

\author{Daisuke Murakami  \\
	Department of Data Science, Institute of Statistical Mathematics, \\
    10-3 Midoricho, Tokyo 190-8562 Japan, Japan \\
    E-mail: dmuraka@ism.ac.jp\\
	\and 
	Gareth W. Peters \\
	School of Mathematical and Computer Sciences, Heriot-Watt \\
     University, Edinburgh, Scotland EH14 4AS UK \\
	\and 
	Tomoko Matsui \\
	Department of Statistical Modeling, Institute of Statistical  \\
    Mathematics, 10-3 Midoricho, Tokyo 190-8562 Japan \\
	\and 
	Yoshiki Yamagata\\
    Center for Global Environmental Research, National Institute for \\
    Environmental Studies, 16-2 Onogawa, Ibaraki 305-8506 Japan \\
	}

\begin{document}
\maketitle


\begin{abstract}
The statistical quantification of temperature processes for the analysis of urban heat island (UHI) effects and local heat-waves is an increasingly important application domain in smart city dynamic modelling. This leads to the increased importance of real-time heatwave risk management on a fine grained spatial resolution. This study attempts to analyze and develop new methods for modelling the spatio-temporal behavior of ground temperatures. The developed models consider higher order stochastic spatial properties such as skewness and kurtosis, which are key components for understanding and describing local temperature fluctuations and UHI's. The developed models are applied to the greater Tokyo metropolitan area for a detailed real world data case study. The analysis also demonstrates how to statistically incorporate a variety of real data sets. This includes remote sensed imagery and a variety of ground based monitoring site data to build models linking city and urban covariates to air temperature. The air temperature models are then used to capture high resolution spatial emulator outputs for ground surface temperature modelling. The main class of processes studied include the Tukey g-and-h processes for capturing spatial and temporal aspects of heat processes in urban environments.
\end{abstract}



\section{Motivation}
\label{sec:1}
Since the first papers on urban heat islands (UHIs) were published (see \cite{oke1982energetic}), the significance of the quantification and modelling of spatially fine granular temperature processes in urban and city environments has grown in prominence. Increasingly it is becoming possible to improve both the resolution of such spatial-temporal temperature models as well as to distinguish between air temperature models and surface/ground temperature models. In the framework developed in this manuscript we will develop novel methods to combine different data sources in order to model both air and surface temperature in local urban environments in a consistent manner.

We see two main directions for the use of such UHI air and ground temperature models. The first is related to climate initiatives and smart city designs. The second is related to policy making on risk reduction and mitigation of population morbidity and mortality effects arising from urban heatwaves, see for instance \cite{semenza1996heat}. We will briefly provide context for these two application domains below.

From the perspective of climate initiatives and smart city design, global initiatives are being established to study such local temperature processes, driven through channels such as green finance by international organizations such as the Climate Bond Initiative. These organizations are actively developing guidance and frameworks for the quantification of pollution reduction in smart city environments. In particular, they are interested in inter-relationships between CO2 emissions in urban environments and local temperature profiles, especially how such temperature profiles can relate to emissions in urban environments( see discussions in \cite{glaeser2010greenness,mathews2010mobilizing,buchner2011landscape}). This interest is driven by a certification and regulation perspective, where funding for low carbon buildings and efficient new transportation systems is increasingly linked to verifiable emissions targets. In such applications there is a strong demand for accurate models of both ground and air temperature spatial-temporal dynamics.  

The importance of local urban temperature modelling can also inform policy for establishing emission reduction standards. Often components of such models are increasingly being used as benchmark references to set standards that should be achieved by green infrastructure projects for transportation and low carbon building initiatives in smart city environments. They are even increasingly being linked to justify access to proceeds from, for instance, green bond debt instruments. The eligibility criteria for such loans is increasingly reliant on scientific certification of the reductions achieved. Quantification of such reductions can be obtained from statistical modelling of local temperature profiles, such as via models developed in this manuscript. The significance of such finance for green project funding by debt instruments, such as green bonds and their associated securitization markets, are projected to globally reach 1-2 trillion USD in the next couple of years. This places an even stronger spotlight on the methods utilized to quantify spatial temperature processes to understand UHIs accurately in order to assess the suitability of environmental projects funded by such debt capital(please see discussions in \cite{kidneygreening}).

In the context of developing local temperature models for health related policy, such local UHI and heat-wave models for both air and ground temperature can help inform policy for mitigation and severity/exposure reduction for at risk populations. Such models can inform strategies to help reduce the high morbidity and mortality rates due to extreme temperature distress, especially in elderly urban populations. To understand the impact such events can have on populations, we note that, during the Great European heatwave in 2003, the number of deaths from heat related distress was estimated to be $>$ 52,000 people. Other similar events include the Russian heatwave in 2010, which had an estimated number of deaths of 15,000 people. Numerous other heatwave events in India, Pakistan, the Middle East, and Australia have also resulted in significant recordings of extremely high morbidity and mortality. As a result of the severity of such heatwave events, from a political perspective, strengthening resilience and adaptability to climate-related hazards, including heatwaves, was selected as one of the 17 agendas of the Social Development Goals (SDGs), which were initiated by the United Nation in 2015. Thus, heatwave monitoring and management is an important issue in smart city design and policy agendas for aging urban populations around the world. 

In this study, we develop application studies on a target area corresponding the Tokyo metropolitan area, which is the largest metropolitan area in terms of population (35 million people). It also contains one of the largest urban/city dwelling elderly populations on the planet. This study attempts to analyze the spatio-temporal behavior of ground temperatures in the Tokyo metropolitan area, focusing on not only mean and variance, which have been considered in many spatial and temporal temperature studies, such as \cite{stewart2012local,murakami2016participatory}, but also skewness and kurtosis, which are key parameters describing extreme heat. 

\section{Statistical Modeling Context}
\label{sec:1.2}
A statistical framework is developed to produce an emulator output in the form of a spatially resolved gridded data set of ground temperature. This emulator output is constructed by combining temperature observations in space and time from a combination of remote satellite sensing imagery, as well as a variety of ground based monitoring site data. The resulting emulator output data is then modelled using Tukey g-and-h regression processes for capturing spatial and temporal aspects of heat processes in urban environments. These processes are the natural extension of Gaussian Process regression models, to incorporate transformations or warping properties that explicitly allow one to parameterise skewess and kurtosis. Spatial modelling is achieved by applying the Tukey g-and-h random fields model to the combined dataset, see \cite{jorge1984some,SaiGWPIN,chen2017dynamic} and \cite{xu2017tukey}. 

As fine-scale spatial and temporal resolutions are utilized, the dataset studied is very large, resulting in a significant computational statistical challenge for the model estimation component of the application. This is reduced by specialized spectral covariance rank reduction when the model is applied to the emulator model ground temperature dataset. 

The resulting analysis of the Tokyo case study suggests that the mean, variance, and spatial skewness are important components in characterizing ground temperatures and subsequent UHI and local heat-wave propensity. It is also observed that model parameters considerably vary between urban, suburban, and mountain areas, and one can cluster such regions to form adaptive local climate policy responses to UHI's in different regions of the urban cityscape.

\subsection{Spatial-Temporal Temperature Modelling for Urban Environments}
\label{sec:1.22}
Spatial and temporal data for aspects of temperature, noise, and pollution are increasingly being collected as part of smart city environmental monitoring initiatives( see \cite{batty2013big,kitchin2014real}). This has been facilitated by developments of remote sensor technology and the global positioning system (GPS), along with other remote sensing modes(see \cite{hancke2012role}). 

In this work, we are concerned with the modelling of such local spatial and temporally resolved data. It is common to develop models for such data via Gaussian process (GP) techniques. However, in the case of fine resolution models applicable to the study of local urban environments, the number of observation points or design points $n$ can be very high dimensional in both time resolution and space resolution. In such cases, the classical estimation of a GP or kriging-based regression framework will require inversion of an $nT \times nT$ covariance matrix, where $n$ is the number of sample sites and $T$ is the number of observation times. The computational cost involved in the estimation and prediction is then of the order $O([nT]^3)$, which is only feasible when $n$ and $T$ are small to medium size. However, when modelling local climate regions, $n$ is typically very large to accommodate all sampling sites. The time resolution is also often on scale of minutes which can make $T$ also very large when combining ground based sensor data with daily satellite data over many months to years.

To address such computational issues, there exists a variety of GP approximation approaches for large sample situations. They include low rank approaches (e.g., fixed rank kriging \cite{cressie2008fixed}, predictive process modeling \cite{banerjee2008gaussian}, and multi-resolution GP \cite{nychka2015multiresolution}), sparse approaches (e.g., covariance tapering \cite{furrer2006covariance}, the local approximate GP \cite{gramacy2015local}, the nearest neighbor GP \cite{datta2016hierarchical} and spatial partitioning approach \cite{kim2005analyzing}, please see \cite{heaton2019case} for a review of approximation approaches. Many such processes have also been extended for spatio-temporal modeling. For example,\cite{banerjee2008gaussian}, \cite{cressie2010fixed}, and \cite{cressie2015statistics} have studied low rank spatio-temporal models; \cite{genton2007separable} and \cite{cameletti2013spatio} have studied sparse spatio-temporal models.

The popularity of such GP and krigin models is largely due to the relative simplicity they provide in model formulation and the statistical characteristics related to the sufficiency of the analysis of second order process information on the spatial-temporal mean and covariance structure of the underlying process. We argue that, in the context of local resolution models, one may wish to develop models to account for higher order stochastic structures that may vary in space and time. In the local climate modelling setting we study, we have large samples in space and time that we believe could potentially have rich information on not just second order process characteristics of mean/variance. Increased sample resolution can also potentially allow for the measurement and therefore modelling of higher order aspects of such spatial temporal processes, such as spatial skewness and kurtosis describing extremal tail behavior. Such higher order information should be included in spatial-temporal models to ensure that resulting estimations are accurate, especially when seeking to explain extreme events such as extreme heat and UHIs. 

Modeling spatial and temporal extremes has recently become a more focused topic in geostatistics owing to increasing availability of data at improved sampling resolutions. The max-stable process \cite{de2007extreme}, which is a natural infinite-dimensional extension of the univariate generalized extreme value distribution, has been extended in a spatial setting, e.g., by \cite{smith1990max}, \cite{kabluchko2009spectral}, and \cite{reich2012hierarchical}.  Their proposed spatial max-stable processes have frequently been applied to model extreme climate events, e.g., by \cite{buishand2008spatial} and \cite{davison2012statistical}. Unfortunately, as the likelihoods associated with the spatial max-stable process have no closed-from expressions (\cite{reich2012hierarchical}), they are not easily developed, especially for spatial-temporal contexts with high resolution sampling, such as when $n$ or $T$ is large. There is also little to no work on the temporal time series characterization of such processes.

Furthermore, such Max-Stable processes assume that the index of regular variation characterizing the spatial extreme behavior is homogeneous in space and time, which is not an assumption we wish to make with the fine resolution process modelling undertaken in this work. Furthermore, such approaches are more applicable to the characterization of only extreme behaviors as opposed to the objective we seek, which is to extend from second order characteristics to higher order stochastic structures while also allowing for regular variation properties of models to capture potential heavy tailed extremes should they appear locally in some regions of space over time.

Therefore, we seek to develop new classes of processes. In this regard, we will build on the recent studies of \cite{xu2017tukey}, \cite{chen2017dynamic} and \cite{nagarajan2018spatial}. In these studies, the different variations in the Tukey G-and H (TGH) random fields (TGH-RF) were proposed. Such models are typically developed as transforms or warpings of a Gaussian process via the Tukey G-and-H transformation. This family of transformations enables explicit modelling and estimation of higher order features such as co-skewness and co-kurtosis of an underlying process(please see explicit derivations of such population process characteristics in \cite{nagarajan2018spatial}). In regard to temporal components, such models have recently been extended by \cite{yan2017non} to capture non-Gaussian auto-regressive processes. 

Warped GP models in the Tukey G-and-H family are now more applicable for modelling in such big data contexts as the challenges associated with the evaluation of the likelihood, first outlined in \cite{peters2006bayesian} and \cite{cruz2014fundamental}, have been resolved in efficient computational frameworks, such as the approach in \cite{xu2015efficient}, where a numerically efficient approach to maximize an approximate likelihood is developed, at least for small $n$ and $T$ sample settings. In this work, we will propose methods to overcome the numerical cost incurred in the inversion of the $nT \times nT$ covariance matrix, which makes computation intractable when $n$ or $T$ are large. Therefore, one of the contributions we present to such modelling frameworks in this paper is a framework to accelerate the TGH-RF approach to make it applicable to large $n$ and $T$ high spatial-temporal resolution contexts.

We can then use such models for observation data based on remotely sensed images for temperature data analysis which produce a very large sample size $n$ in space. To give an indication of how large such data is when trying to train such a GH spatial process model, we note that such training data has had sizes that are typically of the order of gigabytes per minute. To facilitate the use of such data in local temperature modelling contexts, we newly develop two approaches, one based on a low rank TGH-RF modeling and the second based on a version of sparse TGH-RF modeling approach, and we compare them with the original TGH-RF approaches in the references above in terms of computational time and estimation accuracy.  Such approaches are natural extensions of the ideas of the laGP framework (\cite{gramacy2015local}) extended to the GH process context. We discuss and explore such a methodological extension.

Another contribution of this study is development of high-resolution dataset of emulated ground temperatures in the Tokyo metropolitan area; the developed TGH-RF models are applied to this dataset for heatwave analysis. There are at least two monitoring systems available to provide temperature data: station monitoring and remote sensing. The former monitors temperature, humidity, and other climate variables by min and by hour which are observed at spatially distributed monitoring stations. The latter monitors ground temperatures in an ultra high spatial resolution per for a particular observation window per day, however the values are often missing in certain spatial regions due to cloudy/rainy days (see, Section \ref{sec:2}). We combine these data for the temperature emulation. For discussions combining multiple spatial datasets, see \cite{gotway2002combining}.

In short, we develop low rank and sparse TGH-RF modeling approaches, and apply it to the emulated temperature dataset. This study is organized as follows. In Section \ref{sec:2}, we develop an emulator for spatially fine ground temperature maps for each day by combining heat-related data from remote satellite sensing data combined with ground-based weather monitoring station air temperature measurements that are available in the study region, Tokyo greater area. Sections \ref{sec:5} develops and applies our approximated fast TGH-RF models to analyze temperature distribution properties, including spatial co-skewness and spatial co-kurtosis, and compare this novel solution to classical approaches. Section \ref{sec:4} further explores distribution properties of the temperatures. Based on the results, Finally, section \ref{sec:6} concludes our discussion.
\begin{figure}[H]
\begin{center}
  \includegraphics[scale=0.65]{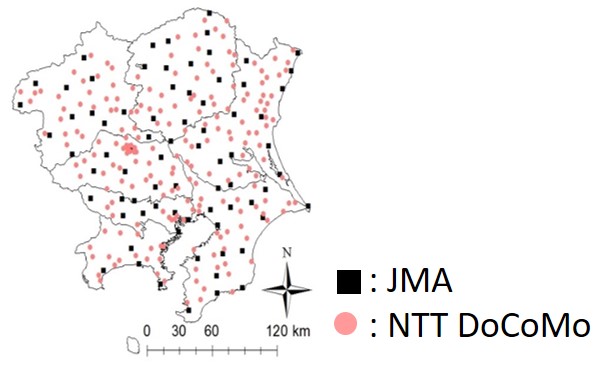}
  \caption{Air temperature monitoring stations in the greater Tokyo area monitored by JMA and NTT DoCoMo.\label{fig:1}}
\end{center}
\end{figure}

\section{Combining Satellite Remote Sensing Data with Ground Monitoring Station Data}
\label{sec:2}
In this section we first outline the properties of the different data sources we consider. Next, we explain the first stage of our analysis, which involves the framework developed to produce emulator model output data that combines satellite remote sensing ground temperature data with ground-based weather monitoring air temperature data as well as spatial covariate information. We need to combine data in an emulator output because the three sources of data we consider have the following attributes:
\begin{itemize}
\item \underline{Ground temperatures:} high spatial resolution and accuracy (Lansat, ASTER and MODIS) satellite data of ground temperatures that is sparse in time and occluded by cloud coverage in some spatial regions on some images;
\item \underline{Air temperatures:} low spatial resolution high accuracy hourly inter-daily temporal resolution ground-based monitoring stations time series data from NTT DoCoMoInc as well as data provided by the Japan Meteorological Ajency (JMA data).;
\end{itemize}

\subsection{Data Characteristics: Air and Ground Temperature (A1, A2)}\label{sec:2.1}
The following air temperature data sets were considered in the target area. The JMA data provides hourly temperatures monitored at 78 monitoring stations (as of 2013) since 1976. The NTTDoCoMoInc data provides air temperatures observed every minute at 206 stations. The data is available from 2013. Figure \ref{fig:1}  shows the plots of the 78 + 206 monitoring stations in our study area.

\begin{figure}
\begin{center}
\includegraphics[scale=0.35]{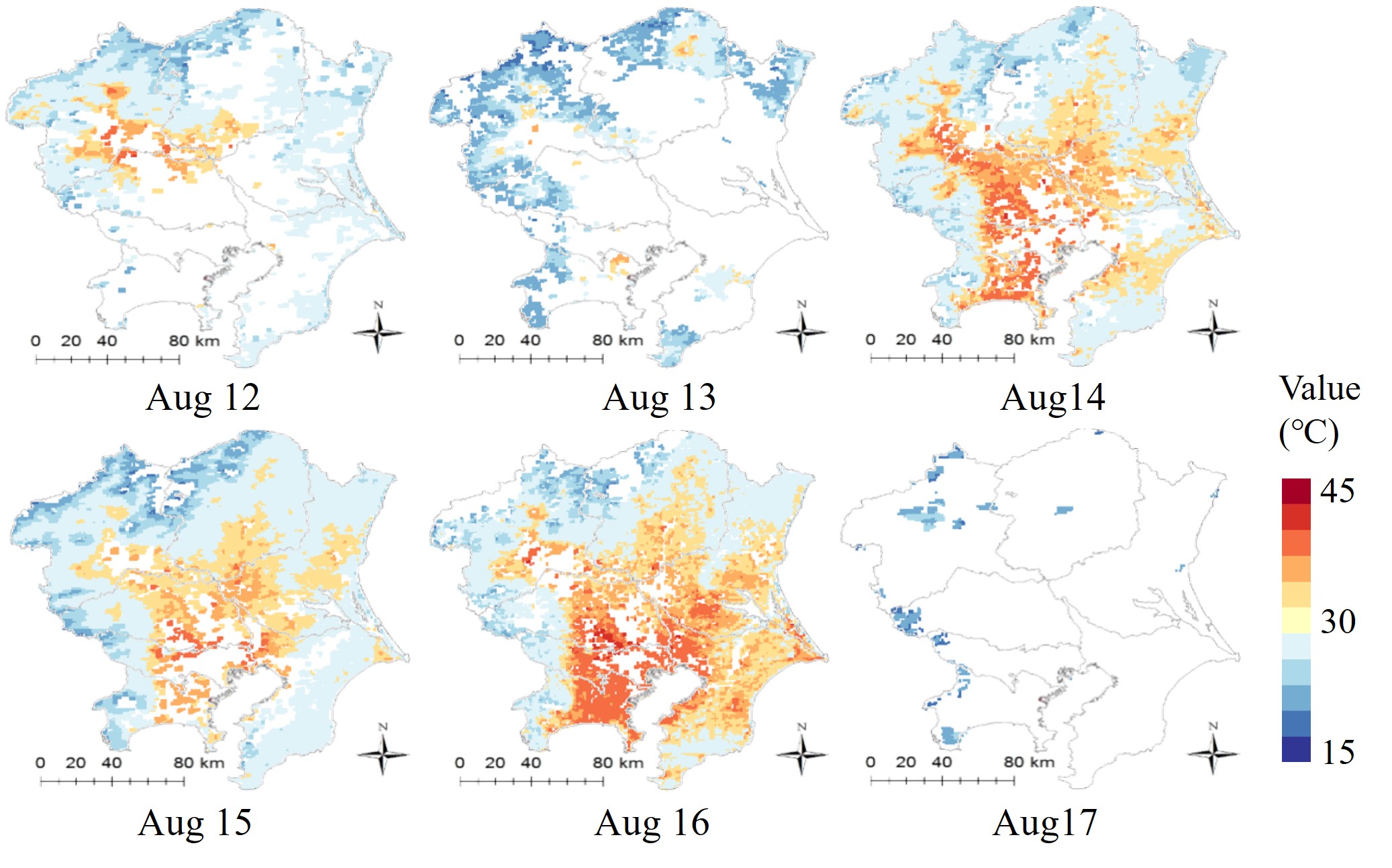}
\caption{Illustrative example of images of MODIS ground temperatures (August 12 - 17, 2013). Blank represents unobserved areas due to obstruction by clouds. \label{fig:2}}
\end{center}
\end{figure}

Regarding ground temperatures, remotely sensed images observed from the MODerate resolution Imaging Spectroradio-meter (MODIS; http://modis.gsfc.\\nasa.gov/) are available. MODIS provides four maps with spatial resolution of 1 km at 10:30, 13:30, 22:30, and 25:30 every day. In the target area, MODIS has 31,235 observation points. We cross checked the records against the other data sources mentioned for a quality assurance. Figure \ref{fig:2} shows the plots of MODIS ground temperatures on 6 days in August, 2013. As you can see, there can be missing values on cloudy/rainy days. Daily mean of the missing ratio in August is 0.70, while the ratio in September is 0.65. Despite the missingness present we were still able to utilize 9,284 and 10,813 samples from the MODIS observatory in August and September, respectively. 

We note that ground temperature is our primary focus as it reflects the absorption and radiation of heat from roads, buildings, and other urban materials, which determines the intensity of heatwaves and UHI's. From the data description, we see that data on both air and ground temperatures are incomplete. The subsequent section explains how to combine these different data sources with local spatial covariate information, which is summarized in Table \ref{table:1}, to develop a regression based emulator model to complete the missing observations in order to obtain a complete ground temperature model output profile in space and time. 

Consideration of these additional location specific explanatory variables was found to be statistically significant in designing our emulator model to capture discontinuous changes in air/ground temperatures around the border separating urban and non-urban areas.

\begin{table*}[!t]
\begin{threeparttable}
 \caption{Explanatory variables \label{table:1}}
    \begin{tabular}{l |lcc}
\hline
Name & Description & Source & Year\\
\hline
Station & 
\begin{tabular}{l}
Exponential of negative distance to \\
the nearest station (accessibility to \\
the nearest railway station) [m] 
\end{tabular}
& NLNI & 2005 \\
\hdashline
Population & Nighttime population density [people/$km^2$] & Census & 2005 \\
\hdashline
Crop land & Proportion of crop land in 1 km grids & & \\
Forest & Proportion of forest land in 1 km grids & & 1997 \\
Urban & Proportion of urban land in 1 km grids & NLNI & 2006 \\
Water & Proportion of water body in 1 km grids &  & 2009 \\
Ocean & Proportion of ocean in 1 km grids & & \\
\hdashline
Elevation & Elevation [m] & NLNI & 2006 \\
\hdashline
Latitude & Latitude [degree]  & JMA  \\
\hline
 \end{tabular}
\begin{tablenotes}
\item[1] 
NLNI: National Land Numerical Information download service \\
(\url{http://nlftp.mlit.go.jp/ksj-e/index.html})
\item[2] 
Census: Population of Japan 2005 (\url{https://www.stat.go.jp/english/\\data/kokusei/2005/poj/mokuji.html})
\item[3] 
JMA: Japan Meteorological Agency \\
(\url{https://www.jma.go.jp/jma/indexe.html})
\end{tablenotes}
\end{threeparttable}
\end{table*}

\subsection{Data Combining Process and Emulator Model Overview}\label{sec:2.2}
\begin{figure}
\begin{center}
\includegraphics[scale=0.45]{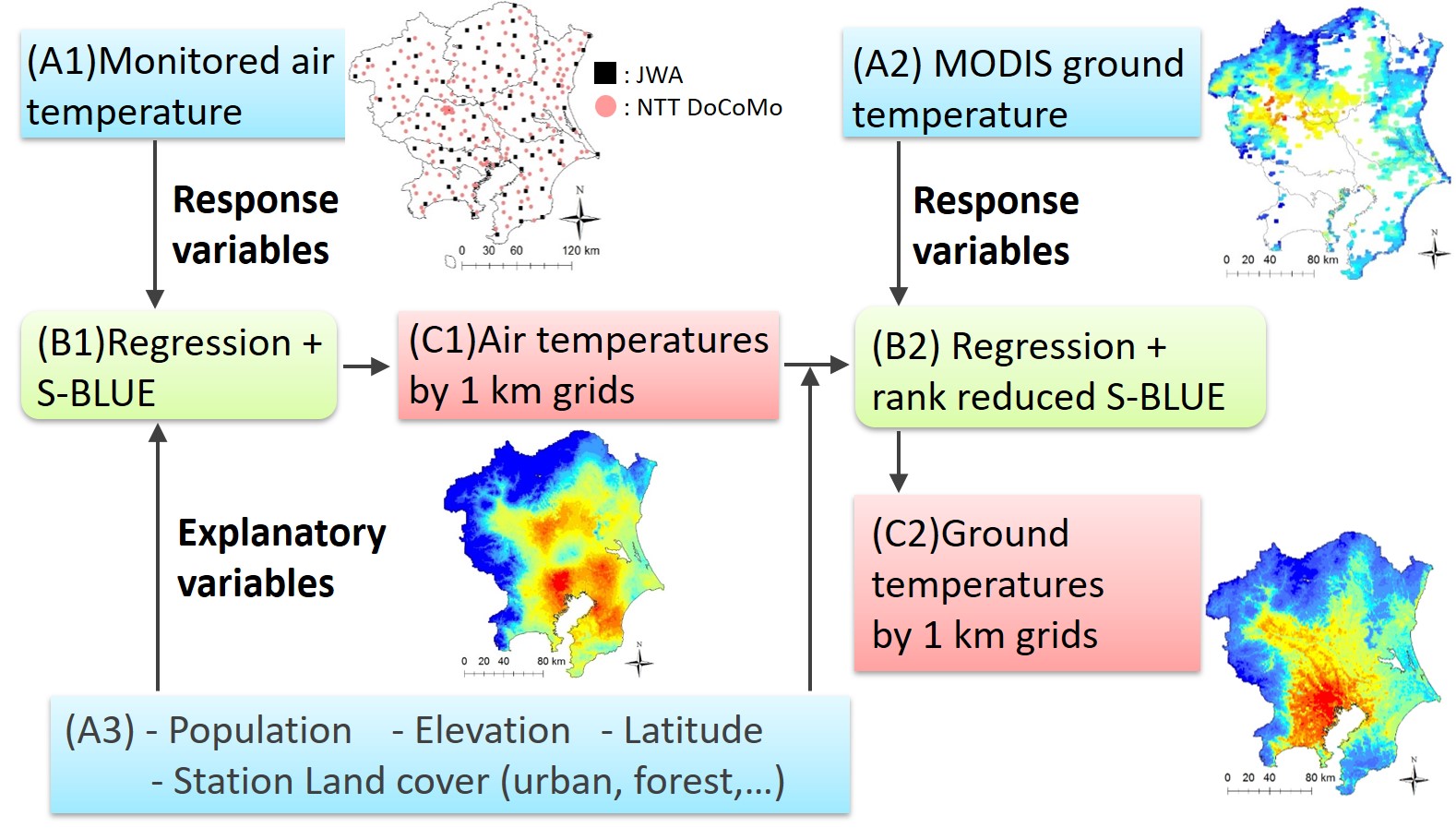}
\caption{Overview of the data processing (Blue: raw input data (A1-3); Green: model fitting (B1-2); Red: emulated data produced from the fitted model (C1-2)). \label{fig:flow}}
\end{center}
\end{figure}

Figure \ref{fig:flow} summarizes the stages involved in the emulator regression model undertaken in order to produce an emulator spatial-temporal map for missing ground temperatures. The input data are (A1) the air temperature data (JWA and NTT DoCoMo) and (A2) the ground temperature data (MODIS). We first model (A1) the air temperatures using (B1) a regression S-BLUE (Spatial Best Linear Unbiased Estimator (\cite{nevat2013random}); see Section \ref{sec:sblue}), in which the explanatory variables are as summarized in Table \ref{table:1}. Then, we obtain (C1) the emulated air temperature data. The emulated air temperatures and variables in Table \ref{table:1} are used as explanatory variables in (B2) a rank reduced S-BLUE approach (see Section \ref{sec:gmodel}) to emulate the (C2) ground temperature data by 1km grids. 

\subsection{Regression S-BLUE for air temperature emulation (B1)}\label{sec:tmodel}

Section \ref{sec:tmodel0} explains the model and Section \ref{sec:sblue} explains the regression S-BLUE approach.

\subsubsection{Spatial regression model}\label{sec:tmodel0}

We assume the following regression structure to model the observed air temperatures:
\begin{align} \label{eq:1}
{\bm Y} = {\bm X} {\bm z} + {\bm \epsilon} \hspace{1cm} E[{\bm \epsilon}]={\bm 0} \hspace{1cm} Cov[{\bm \epsilon}] ={\bm C},
\end{align}
where ${\bm Y}= \left[ Y(s_1),\ldots , Y(s_i),\ldots , Y(s_n) \right]^\mathrm{T}$ is a vector of air temperatures monitored at locations $ \{ s_1,\ldots s_i, \ldots s_n \} \in D \subseteq \mathbb{R}^2$, ${\bm X}$ is a $(n \times K)$ matrix of the explanatory variables summarized in Table \ref{table:1}, ${\bf z}$ is a vector of regression coefficients, and $ {\bm \epsilon}$ represents the noise process.

The $(n \times n)$ covariance matrix of the noise process, denoted ${\bf C}$, is parametrized by a distance-decay exponential kernel capturing spatial dependence:
\begin{equation} \label{eq:exp}
c[d(s_i, s_j)]= 
\begin{cases}
\tau^2 exp \Bigl( - \frac{d(s_i, s_j)}{r} \Bigr) & if \hspace{0.5cm}  d(s_i, s_j) > 0 \\
\tau^2 + \sigma ^ 2 & otherwise,
\end{cases}
\end{equation}
where $d(s_i, s_j)$ is the Euclidean distance between sites $s_i$ and $s_j$ measured in units of meters [m], $r$ is a range parameter, and $\tau^2$ and $\sigma^2$ are variance parameters for spatial and non-spatial variations, respectively. The kernel can be replaced with Gaussian, spherical, or other positive definite kernels (see \cite{cressie2015statistics}). Our model describes discontinuous change of temperatures, e.g., at borders separating urban and non-urban areas, using the regression term while continuous change using the spatially dependent process. 

The same structure is assumed behind unobserved temperature at location $s_\ast  \in D \subset \Re^2$.
\begin{align} \label{eq:2}
Y(s_\ast) = {\bm X}(s_\ast)^\mathrm{T} {\bf z} +\epsilon(s_\ast)  &{  }& Cov[\epsilon(s_\ast), {\bm \epsilon}] ={\bm c}(s_\ast),
\end{align}
where ${\bm X}(s_\ast)$($K \times 1$) are the explanatory variables observed at the site $s_\ast$, ${\bm c}(s_\ast)$ ($n \times 1$) is a vector whose $i$-th element equals $c[d(s_\ast, s_i)]$.

\subsubsection{Liner approximation by using S-BLUE}\label{sec:sblue}

This study linearly approximates a non-linear spatial smoothing function behind air temperatures using a spatial best linear unbiased estimator, termed the S-BLUE, as studied in \cite{nevat2013random}. S-BLUE, which we denote $\hat{f}(s_\ast)$, is a linear approximation of a non-linear spatial smoothing function to estimate the missing observation $Y(s_\ast)$ by minimizing the mean squared error $E \left[ (y(s_\ast)-\hat{f}(s_\ast))^2 \right]$ under the constraints of linearity and unbiasedness. By solving the minimization problem under Eqs. (\ref{eq:1}, \ref{eq:2}), the regression S-BLUE estimator is given as follows:
\begin{equation} \label{eq:6}
\hat{f}(s_\ast)={\bm X}(s_\ast)^\mathrm{T} \hat{\bm z} + \hat{\bm c}(s_\ast)^\mathrm{T} \hat{\bm C}^{-1}[{\bm Y}-{\bm X}\hat{\bm z}]
\end{equation}
where $\hat{\bm z}=({\bm X}^\mathrm{T}\hat{\bm C}^{-1}{\bm X})^{-1}{\bm X}^\mathrm{T}\hat{\bm C}^{-1}{\bm Y}$. We use this regression S-BLUE to emulate air temperatures by 1 km grids. The variance parameters $\{ \tau^2, \sigma ^ 2, r \}$ are estimated by the robust weighted least squares (WLS) method of \cite{cressie1980robust}. We used an R package of gstat (https:\slash\slash{}cran.r-project.org \slash{}web\slash{}packages\slash{}gstat\slash) for the estimation.

\subsection{Emulated ground temperatures (C1)}\label{sec:atemp}

We used the (B1) the regression S-BLUE approach to emulate air temperatures by 1 km grid at 13:00 JST each day for 61 days. For days with less than 5,000 observations we use data on the day and one day before and after to stabilize the estimation. We illustrate a randomly selected, but representative example of the fit results below.

\begin{figure}
\begin{center}
\includegraphics[scale=0.35]{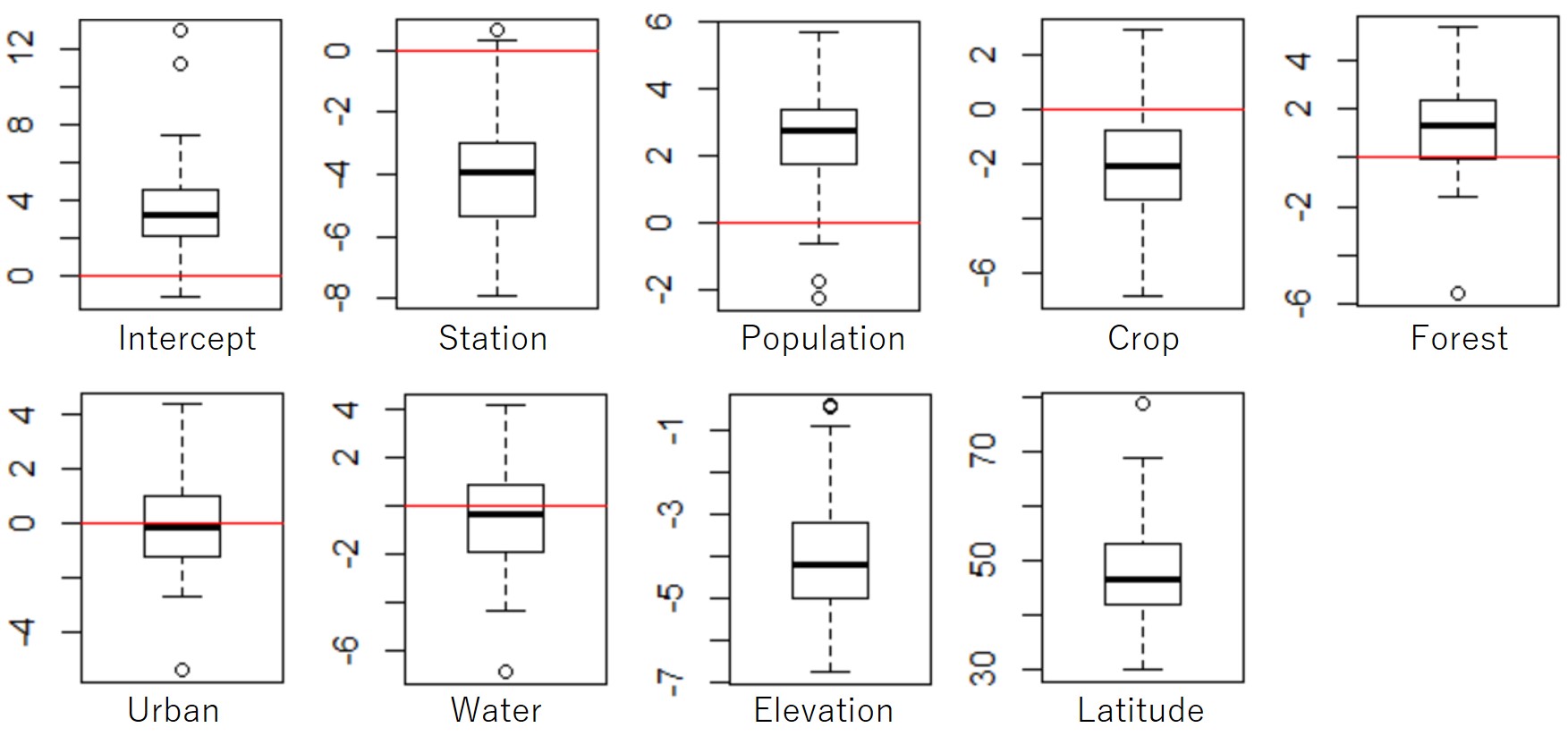}
\caption{Boxplots of the estimated t-values of the regression coefficients in each day (air temperature model) \label{fig:tval_a}}
\end{center}
\end{figure}

Figure \ref{fig:tval_a} summarizes boxplots of the estimated t-values of the regression coefficients for the air temperature model. Note that the variance inflation factor (VIF), which exceeds 10 if there is a severe multicollinearity, verifies that the explanatory variables are not collinear with each other. It can be seen that the air temperatures are significantly low in crop land areas while significantly high in populated area. These results are plausible considering the cooling effects of natural environments and the trapping of heat due to the UHI effects in populated areas. Furthermore, we see that elevation is negatively significant. The positive sign of Latitude might reflect the ocean's cooling effect in the south area while the thermal storage effect of inland areas in the north area.

\begin{figure}
\begin{center}
\includegraphics[scale=0.3]{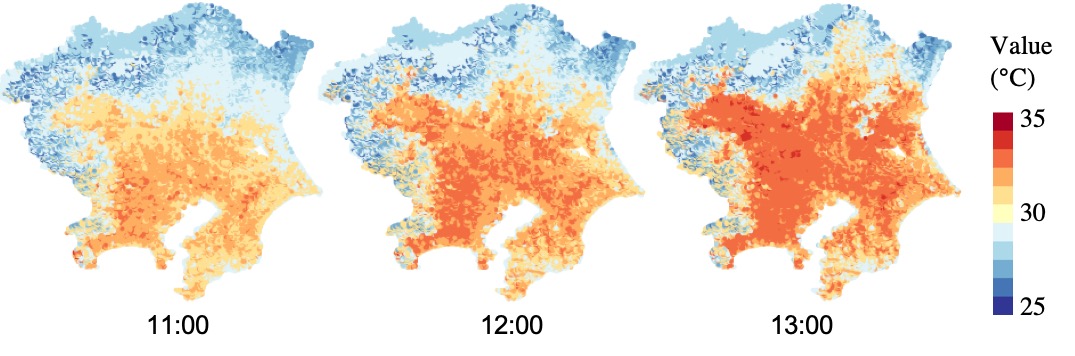}
\caption{Emulated air temperatures in the Tokyo metropolitan area at 11:00, 12:00, and 13:00 on August 17. \label{fig:atemp}}
\end{center}
\end{figure}

Air temperatures emulated at 11:00, 12:00, and 13:00 on August 15 are plotted in Figure \ref{fig:atemp}. These panels show a rapid increase in air temperatures, especially near the urban area.

\subsection{(B2) Rank reduced regression S-BLUE for ground temperature emulation}\label{sec:gmodel}

The variables summarized in Table \ref{table:1} and (C1) the emulated air temperatures are used as explanatory variables for (B2) the ground temperature emulation. Unfortunately, the original S-BLUE, which has $\hat{\bm C}^{-1}$, is computationally infeasible because of the large sample size of the MODIS data. To lighten the cost, we approximates $\hat{\bm C}$ as $\hat{\bm  E}_L\hat{\bm \Lambda}_L\hat{\bm E}_L^\mathrm{T}$, where $\hat{\bm \Lambda}_L$ is an $L \times L$ diagonal matrix that consists of the $L$ largest eigenvalues approximated by the Nystr$\ddot{\rm o}$m extension (\cite{williams2001using}), and $\hat{\bm  E}_L$ is an $n \times L$ matrix that consists of the corresponding $L$ approximated eigenvectors Thus, we approximate the $\hat{\bm C}$ using the $L$ principal eigen-pairs. Given the eigen-decomposition, the S-BLUE estimator yields:
\begin{equation} \label{eq:6}
\hat{f}(s_\ast)={\bm X}(s_\ast)^\mathrm{T} \hat{\bm z} + \hat{\bm c}(s_\ast)^\mathrm{T} \hat{\bm  E}_L\hat{\bm \Lambda}_L^{-1} \hat{\bm E}_L^\mathrm{T}[{\bm Y}-{\bm X}\hat{\bm z}]
\end{equation}
The complexity for the rank reduced regression S-BLUE is $O(n)$ that is feasible for the MODIS data. 

\subsection{(C2)Emulated ground temperatures}\label{sec:gtemp}

\begin{figure}
\begin{center}
\includegraphics[scale=0.35]{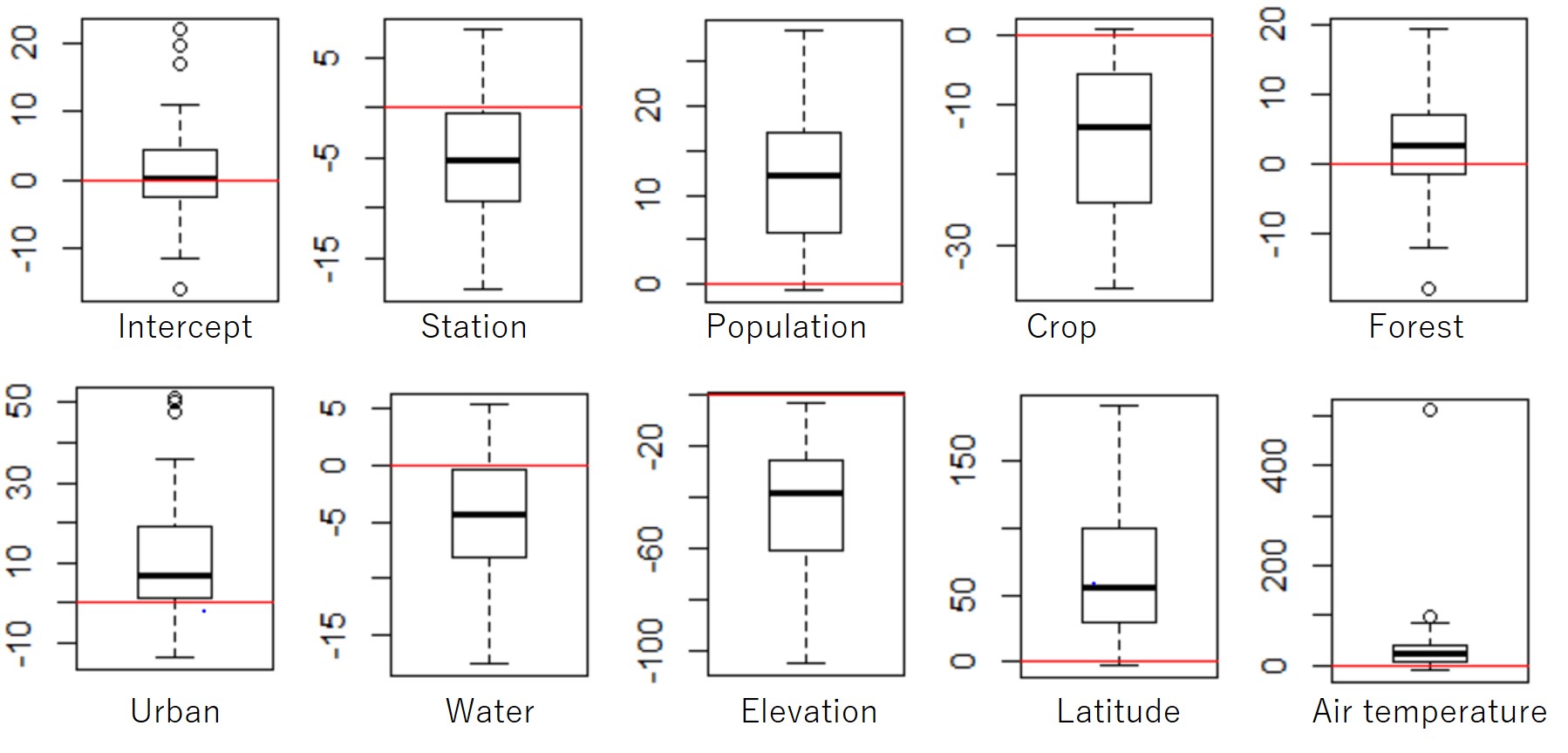}
\caption{Boxplots of the estimated t-values of the regression coefficients in each day (ground temperature model). \label{fig:tval_g}}
\end{center}
\end{figure}

Ground temperatures by 1 km grids 13:00 JST each day for 61 days are emulated using the rank reduced S-BLUE approach. Based on a preliminary analysis, $L$ is given by 200. Here, to linearize the relationship between ground temperatures and air temperatures, we apply the box-cox transformation to the air temperatures in which the multiplier is estimated a priori by maximizing the likelihood of the linear model between these two.
Figure \ref{fig:tval_g} displays the estimated t-values of regression coefficients. As expected, air temperatures increase ground temperatures. The population density and urban areas have positively significant effect on the ground temperature. These results confirme the existance of UHI  in this area. Furthermore, we find that latitude is positively significant in the model, which can be attributed to that fact that there is a basin that is known to be hot in the northern area of the region of study in the greater Tokyo Metropolitan area.

\begin{figure}
\begin{center}
\includegraphics[scale=0.3]{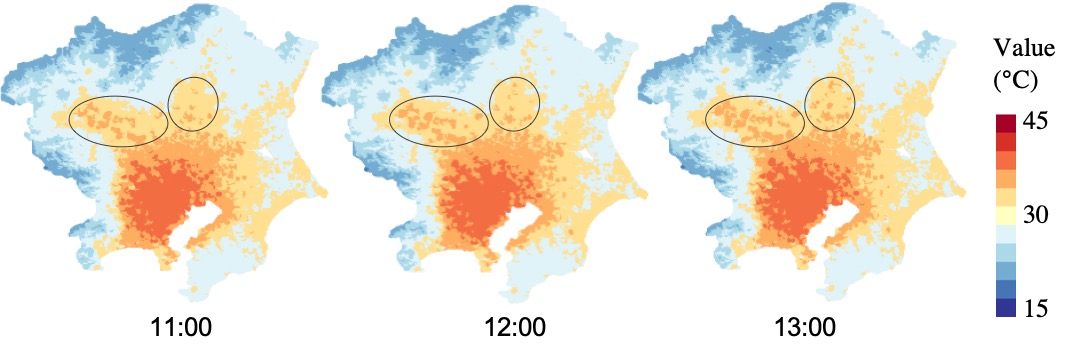}
\caption{Emulated ground temperatures at 11:00, 12:00, 1and 3:00 on  August 17. \label{fig:gtemp}}
\end{center}
\end{figure}

Ground temperatures interpolated at 11:00, 12:00, and 13:00 on August 17 are plotted in Figure \ref{fig:gtemp}. The emulated ground temperatures have have different spatial patterns with air temperatures; ground temperature captures heat in the central area more clearly. This is because ground temperature reflects thermal storage and radiation due to urban materials that are principal sources of urban heat island (e.g., \cite{okada2013proposal}). This is also the reason why ground temperatures are greater than air temperatures. Although air temperatures are usually used for heatwave risk assessment (e.g., \cite{kim2016projection}; \cite{lin2017climate}), our result suggests that ground temperature can be a better indicator for urban heat island. Note that, although the results are only for one of the 61 target days, we confirmed that ground temperatures are high in urban areas on most of the target days. Although temperature increase is conceivable in suburban areas shown in circles, temperature increase over time is relatively small because heats are already saturated at 11:00. 

\begin{figure}
\begin{center}
\includegraphics[scale=0.35]{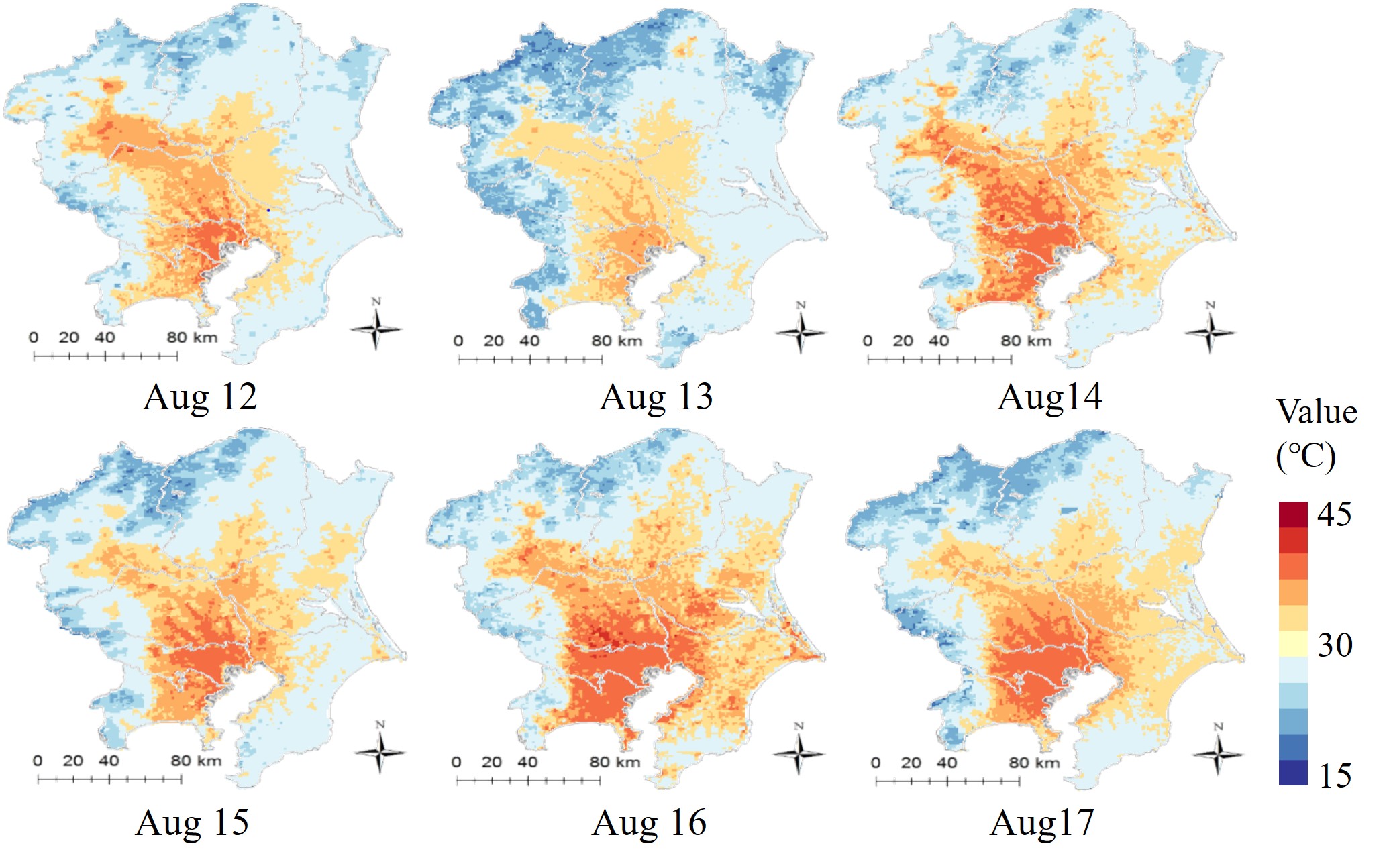}
\caption{Emulated ground temperatures (August 12 - 17). \label{fig:interp}}
\end{center}
\end{figure}

Lastly, emulated ground temperatures at 13:30 from August 12 to 17 are plotted in Figure \ref{fig:interp} for comparison with the original observations (Figure \ref{fig:2}). The result confirms that the emulated temperatures are reasonably high in urban areas while low in mountain areas even in August 17 when missing observations are dominant.

The emulated ground temperature data is available from (https:\slash\slash{}figshare.com\slash{}s\slash{}5a7e2d792d5968078cc7).

\section{Geostatistical analysis of the emulated ground temperatures}\label{sec:5}
\subsection{Outline}\label{sec:5_1}

The previous section emulated daytime ground temperatures by 1 km grids by 61 days. The ground temperatures are represented by $\tilde{Y}_t(s_i)$, where $i \in \{ 1, \ldots, n \}$ and $t \in \{ 1, \ldots, 61 \}$ indicate the grids and days, respectively. As a preliminary study (see Supporting Material A), we analyzed distribution property of the ground temperatures. The result demonstrates the existence of skewness and kurtosis in temperature distribution. It is also suggested that strength of these two considerably change depending on location. 

Given that, this section analyzes the process behind ground temperatures considering skewness and kurtosis using the Tukey g-and-h random fields (TGH-RF) model. After introducing the classical Tukey g-and-h (TGH) distribution in Section \ref{sec:5_1b}, Section \ref{sec:5_2} introduces the TGH-RF model. Unfortunately, this model is computationally intensive and not suitable for our analysis. Accordingly, we extended the model to a low rank model, which estimates global $\{a, b, g, h \}$ values, which are assumed constant over space, in Section \ref{sec:5_3} and a sparse model, which estimates local $\{a, b, g, h \}$ values in Section \ref{sec:5_4}. Then, the models were applied to ground temperature modeling in each of these sections. In each modeling, the four parameters are estimated in each day individually,

\subsection{Tukey g-and-h distribution}\label{sec:5_1b}

The Tukey G-and-H (TGH) distribution is formulated as follows:
\begin{eqnarray}
\tilde{Y} &=&  a + b \tau_{g, h}[Z]  \label{eq:gh1},\\
\tau_{g, h}[Z] &=& \frac{1}{g} {exp \Bigl(g Z - 1 \Bigr)} exp \Bigl(h \frac{Z^2}{2} \Bigr),
\end{eqnarray}
where $Z \sim N(0, 1)$. $\{ a, b, g, h \}$ represent the mean, standard deviation (scale), skewness, and kurtosis of the temperature process. The TGH distribution is useful to analyze skewness and kurtosis. The next section introduces the TGH-RF model combining the TGH distribution with GP to consider spatial dependence.

\subsection{Tukey g-and-h random fields (TGH-RF) models}\label{sec:5_2}
\subsubsection{Original model}\label{sec:5_2_1}

The TGH-RF \cite{xu2016tukey} is an extension of the GP model incorporating skewness and kurtosis parameters. The TGH-RF model is formulated as follows:
\begin{equation} \label{eq:gh_rf1}
\begin{split}
\tilde{Y}_t(s_i) =  a + b \tau_{g, h}[Z_t(s_i)]
\end{split}
\end{equation}
\begin{equation} \label{eq:gh_rf2}
\begin{split}
\tau_{g, h}[Z_t(s_i)] = \frac{1}{g} {exp \Bigl(g Z_t(s_i) - 1 \Bigr)} exp \Bigl(h \frac{Z^2_t(s_i)}{2} \Bigr),
\end{split}
\end{equation}
$Z_t(s_i)$ is a Gaussian random variable satisfying $E[Z_t(s_i)] = 0$, $Var[Z_t(s_i)] = 1$, $Cov[Z_t(s_i), Z_t(s_j)] = c[d(s_i, s_j)]$. The TGH-RF model is an extention of the Gaussian process model to estimate skewness ($g$) and kurtosis ($h$).

The log-likelihood of the model is given as
\begin{eqnarray} \label{eq:lik1}
\begin{split}
L({\bm \theta}_1, {\bm \theta}_2) \propto &-\frac{1}{2} {\bm Z}_t^\mathrm{T} \Bigl( {\bm C}^{-1} + h {\bm I} \Bigr) {\bm Z}_t -\frac{1}{2} |{\bm C}| - n log(b) \\
& -\sum_{i} log \Bigl[ exp \bigl( g Z(s_i) \bigr) \\
& +\frac{1}{g}{exp \bigl( g Z(s_i)-1 \bigr)} h Z(s_i,) \Bigr]
\end{split}
\end{eqnarray}
where $Z_t(s_i) = \tau_{g,h}^{-1}\Bigl[\frac{\tilde{Y}(s_i)-a}{b}\Bigr]$, \\ ${\bm Z}_t = [Z_t(s_1)$, $Z_t(s_2), \cdots, Z_t(s_n)]$, and ${\bm C}$ is a spatial correlation matrix whose ($s_i$, $s_j$)-th element is $c[d(s_i, s_j)]$. 

${\bm Z}_t$ and $Z(s_i)$ include ${\bm \theta}_1 \in \{a, b, g, h\}$, whereas ${\bf C}$ includes ${\bm \theta}_2$, which are parameters describing spatial dependence. \cite{xu2015efficient} propose the following parameter estimation procedure:
\begin{description}
\item[(i)] Set the initial values for $\hat{\bm \theta}_1$ and $\hat{\bm \theta}_2$.
\item[(ii)] Iterate the following steps until parameter estimates converge.
 \begin{description}
  \item[(ii-1)] Numerically maximize $L({\bm \theta_1}, \hat{\bm \theta_2})$ with respect to ${\bm \theta_1}$
  \item[(ii-2)] Numerically maximize $L(\hat{\bm \theta_1}, {\bm \theta_2})$ with respect to ${\bm \theta_2}$
 \end{description}
\end{description}
Unfortunately, this algorithm, which requires an iterative calculation of ${\bf C}^{-1}$ and $|{\bf C}|$, is not available for our data due to the large sample size. The subsequent two subsections introduce the low rank and sparse approaches.

\subsection{Low rank TGH-RF modeling}\label{sec:5_3}
\subsubsection{Model and estimation}\label{sec:5_3_1}

To accelerate the computation, we approximate ${\bm Z}_t$ with ${\bm E}_L {\bm \Gamma}_L$, where ${\bm \Gamma}_L \sim N(\bm{0}, p {\bm \Lambda}_L^m)$, ${\bm E}_L$ is a $n \times L$ matrix of the first $L$ eigenvector of a pre-specified spatial correlation matrix, ${\bf R}$, ${\bm \Lambda}_L$ is a $L \times L$ diagonal matrix whose elements are the first $L$ eigenvalues of the matrix, and $p = \frac{1}{Tr[{\bm \Lambda}_L^m]}$, which is required to ensure $Var[{\bm E}_L {\bm \Gamma}_L] = 1$. $m$ is a parameter determining the scale of the spatial process. Eigenvectors corresponding to greater eigenvalues are emphasized when $m$ is large, and the resulting GP has a global map pattern; the reverse is true for small $m$. Thus, ${\bm Z}_t$ satisfying  $E[{\bm Z}_t]={\bm 0}$, $Var[{\bm Z}_t] = {\bm I}$, and $Corr[{\bm Z}_t] = {\bm C}$ is approximated by ${\bm E}_L {\bm \Gamma}_L$ satisfying $E[{\bm E}_L {\bm \Gamma}_L] = {\bm 0}$, $Var[{\bm E}_L {\bm \Gamma}_L] = {\bm I}$, and $Corr[{\bm E}_L {\bm \Gamma}_L] = {\bm E}_L {\bm \Lambda}_L {\bm E}_L^\mathrm{T}$, which is a low rank approximation of ${\bf C}$. Note that the cost for the eigen-decomposition, whose computational complexity equals $O(n^3)$, can be reduced by using the Nystr$\ddot{\rm o}$m extension (\cite{williams2001using}), the sparse greedy approximations (\cite{smola2001sparse}), or other eigen-approximation technique (see \cite{peters2017statistical}).

The ${i, j}$-th elements of the ${\bm C}$ matrix are given by the exponential kernel, $exp(-d_{i,j}/r)$, where $d_{i, j}$ is the Euclidean distance between sites $i$ and $j$, and $r$ is a given range parameter. Following \cite{dray2006spatial}, $r$ is given a priori by the maximum distance of the minimum spanning tree covering the sample sites. Although the fixed $r$ is somewhat restrictive, the scale parameter $m$ is known to assume the role of the range parameter (\cite{schabenberger2017statistical}). More importantly, the fixed $r$ drastically accelerates the computation as we will demonstrate below.

The log-likelihood Eq.(\ref{eq:lik1}) can be rewritten using the eigen-pairs, as follows:
\begin{eqnarray} \label{eq:lik2}
\begin{split}
L({\bm \theta_1}, {\bm \theta_2}) \propto &-\frac{1}{2} {\bm Z}_t^\mathrm{T} {\bm E} (p {\bm \Lambda}_L^{-m} + h {\bm I})  {\bm E}_L^\mathrm{T} {\bm Z}_t \\
&-\frac{1}{2} p \prod_{l=1}^L {\bm \Lambda}_l^m  - n log(b)\\
&-\sum_{i} log \Bigl[ exp \bigl( g Z_t(s_i) \bigr) \\
&+ \frac{1}{g} {exp \bigl( g Z_t(s_i) \bigr)-1)} h Z_t(s_i) \Bigr]
\end{split}
\end{eqnarray}
${\bm Z}_t ({\bm C}^{-1} + h {\bm I}) {\bm Z}_t$ and $|{\bm C}|$, whose complexity which was formerly $O(n^3)$, is replaced with ${\bm Z}_t^\mathrm{T} {\bm E} (p {\bm \Lambda}_L^{-m} + h {\bm I}) {\bm E}_L^\mathrm{T} {\bm Z}_t$ and $\prod_{l=1}^L \lambda_l^m$, with complexities $O(Ln)$ and $O(L)$, respectively. Note that the computationally efficient log-likelihood specification is not available without fixing the $r$ parameter (because the eigen-pair changes depends on the $r$ value). Eq. (\ref{eq:lik2}) can be maximized just like Eq.(\ref{eq:lik1}), in which ${\bm \theta}_2 = m$. 

We performed a Monte Carlo simulation experiment to evaluate the approximation accuracy of the low rank approach relative to the original TGH-RF model. This result reveals that the low rank approach accurately approximates the {$a, b, g, h$} parameters in a computationally efficient manner as long as the number of basis functions is not too small. Please see Supporting Material B for details of the validation study.

\subsubsection{Application to the emulator model ground temperatures}\label{sec:5_4}

\begin{figure}
\begin{center}
\includegraphics[scale=0.4]{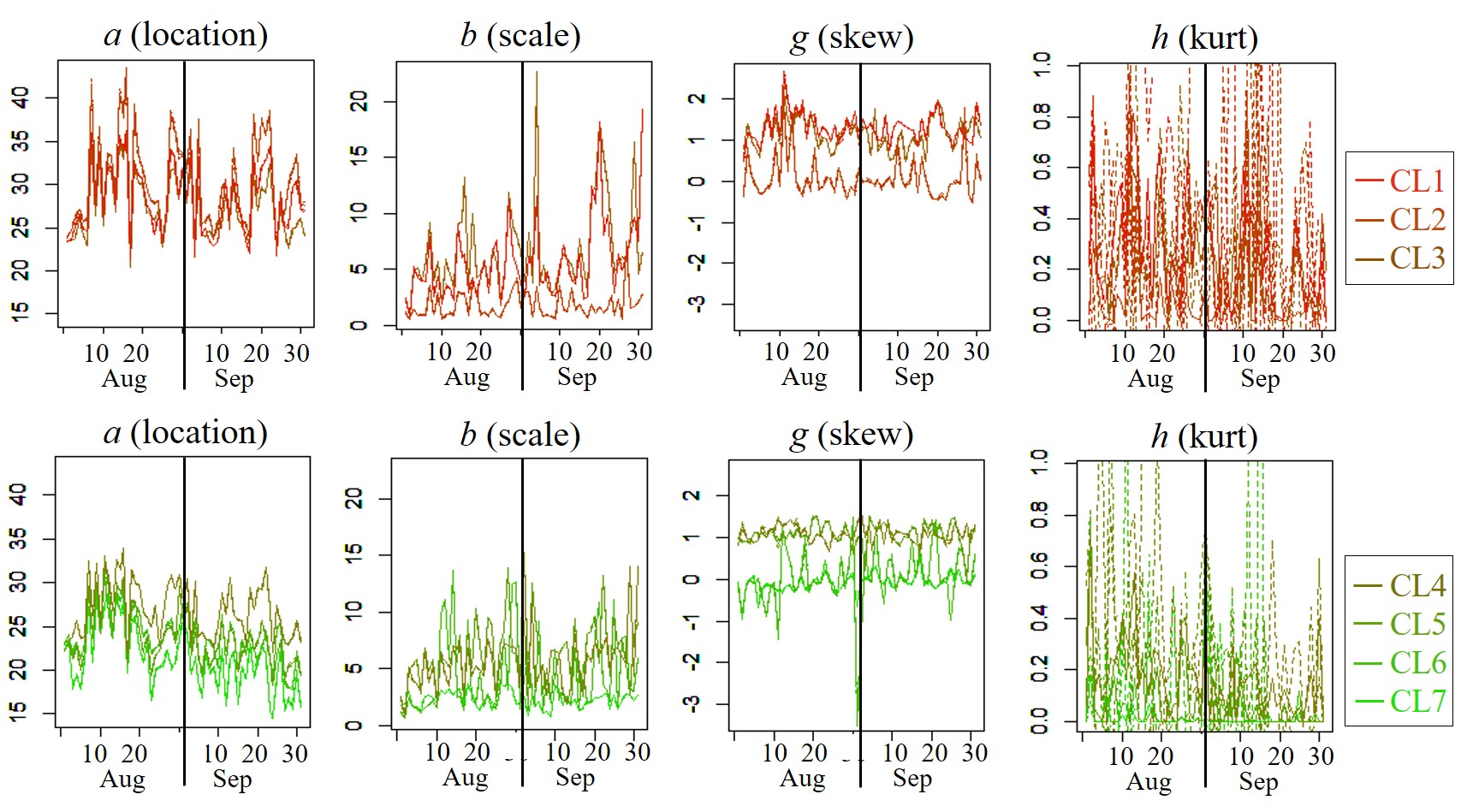}
\caption{Estimated parameters for August and Septemper, 2012. Top: estimated parameters for the three urban clusters whose geometric centers are the closest to the Tokyo station, CL1-3; Bottom: estimated parameters for the other 4 non-urban clusters, CL4-7. In each panel, the solid line represents parameter estimates, and the dash line represents their 95 $\%$ confidential intervals.\label{fig:9}}
\end{center}
\end{figure}

This section applies the low rank approach to estimate the moment parameters in the 10 clusters during the 62 days. Figure \ref{fig:9} summarizes the estimated parameters. The top panels show results in the three urban clusters whose geometric centers are the closest to the Tokyo station, and the bottom shows the other 4 non-urban clusters (see Figure \ref{fig:A2}). In each panel, the solid line represents parameter estimates, and dashed line represents their 95 $\%$ confidential intervals.

In non-urban clusters, the $a$ parameter gradually decreases over the period. By contrast, in urban clusters, large $a$ values last across the period. This may be arising due to the fact that urban materials store heat. Heatwaves are estimated to last longer in urban area than in non-urban areas. The gradual increase in the scale parameter $b$ in urban clusters shows that heat at the end of the summer is more uncertain than at the beginning of the summer.

$g$ values are positively significant in many clusters. The temperatures are estimated to have a right-skewed tail. In other words, the temperatures can take extremely high values, which implies hazardous heat, in rare cases. The importance of considering skewness, which is ignored in standard second-order spatial models (e.g., the Gaussian process model), in heat risk estimation is suggested. By contrast, $h$ values, which have large standard errors, are statistically insignificant in most cases. Kurtosis might be less important in heatwave modeling.

\begin{figure}
\begin{center}
\includegraphics[scale=0.35]{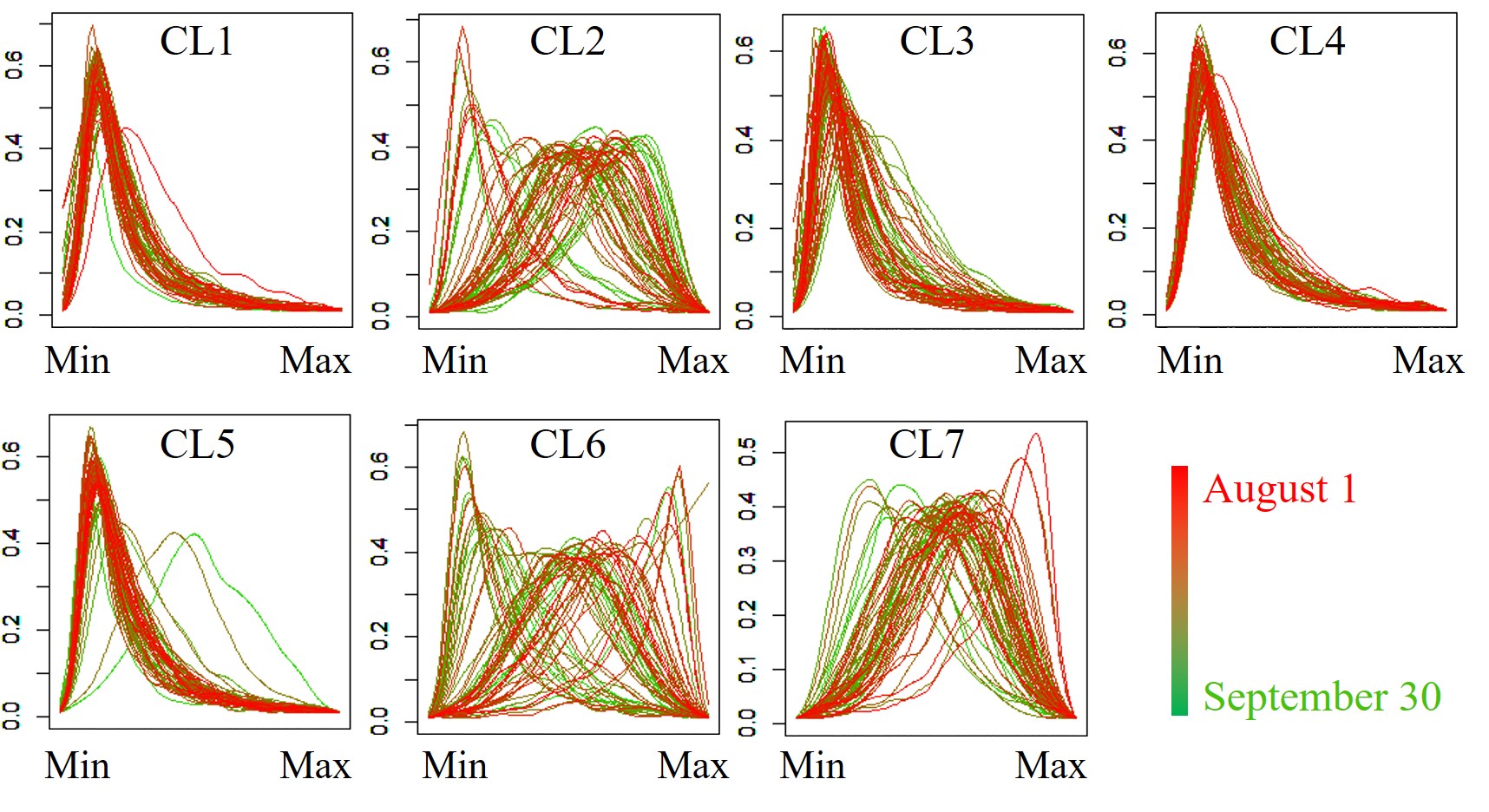}
\caption{Ground temperature distribution for each day and for each cluster. The temperature values are normalized within 0-1 using the maximum and minimum values for each day.\label{fig:10}}
\end{center}
\end{figure}

To study the estimated temperature distributions, the estimated probability densities during each day are drawn in Figure \ref{fig:10}. This figure suggests that the temperature distributions are right-skewed in clusters 1, 3, 4, and 5, which are near the center, and are nearly Gaussian in cluster 6 and 7, which are in the mountain areas. If we assume a Gaussian distribution for temperatures, the right skew is ignored, and risk is underestimated. The usefulness of the low rank TGH-RF model is verified in terms of an accurate heatwave risk estimation.

\subsection{Sparse TGH-RF modeling}\label{sec:5_2_3}
\subsubsection{Outline}\label{sec:5_2_3}

The low rank approach has the following limitations. First, it tends to underestimate small-scale spatial variations (see, \cite{stein2014limitations}). Second, it assumes the same $a, b, g, h$ values across space, although these values might actually vary across geographical space. To overcome these limitations, we developed a sparse TGH-RF approach by extending \cite{gramacy2015local}.

Their approach, which is called local approximate GP (laGP), estimates a local GP for each prediction location independently, using $n(s_\ast)$ local subsamples, which we denote $D(s_i)$. Although the $n(s_i)$ nearest-neighbors (NN) works well in many situations, the NN-based design is not optimal because accurate GP estimation requires some spread of subsamples (\cite{gramacy2016speeding}). Instead,  \cite{gramacy2015local} proposed a greedy search algorithm to find $D(s_\ast)$, minimizing the Bayesian mean-squared predictive error (MSPE). This section accelerates our TGH-RF model by applying their approach.

\subsubsection{Local approximate TGH-RF}\label{sec:laGH}
We assume the following local TGH process around a location $s_0$:  
\begin{equation}\label{eq:laGH}
\begin{split}
\begin{pmatrix}
{\bm y}({\bm s}_{j+1})\\
y(s_0)
\end{pmatrix}
& \sim TGH  [ \tilde{\bm \mu}, \tilde{\bm C}, \tilde{\bm g}, \tilde{\bm h} ] \\
&\tilde{\bm \mu} = 
  \begin{pmatrix}
  {\bm \mu}({\bm s}_{j+1}) \\
  \mu(s_0)
  \end{pmatrix},\\
&\tilde{\bm C}  = \begin{pmatrix}
  {\bm C}({\bm s}_{j+1}, {\bm s}_{j+1}) & {\bm c}({\bm s}_{j+1}, s_0)  \\
  {\bm c}^\mathrm{T}({\bm s}_{j+1}, s_0) & c(s_0, s_0)
  \end{pmatrix},\\
&\tilde{\bm g} = \begin{pmatrix}
    {\bm g}({\bm s}_{j+1}) \\
    g(s_0)
  \end{pmatrix},\\
&\tilde{\bm h} = \begin{pmatrix}
    {\bm h}({\bm s}_{j+1}) \\
    h(s_0)
\end{pmatrix},
\end{split}
\end{equation}
  where ${\bm s}_{j+1}=\{s_1,s_2,\ldots s_{j+1} \}$ is a set of the $j+1$ nearest locations from the location $s_0$ and ${\bm y}({\bm s}_{j+1})$ is a vector of observations at the $j+1$ sites. ${\bm \mu}({\bm s}_{j+1})$  and $\mu(s_0)$ are mean functions. $\tilde{\bm C}$ is a kernel matrix used to model residual spatial dependence. $g(s_0)$ and $h(s_0)$ are the g-and-h parameters around the location $s_0$.
  Given Eq.(\ref{eq:laGH}), we estimate the conditional mean $a(s_0)$ and variance $b(s_0)$ of $y(s_0)$:
    \begin{equation}\label{eq:cond_laGH}
  y(s_0)|{\bm y}({\bm s}_{j+1}) \sim TGH \left[ a(s_0), b(s_0), g(s_0), h(s_0) \right]
  \end{equation}
  
  The expectation of $a(s_0)$ is given by
  \begin{equation}\label{eq:mu}
  \begin{split}
  a(s_0)|{\bm y}({\bm s}_{j+1}) &= \mu(s_0) + {\bm c}^\mathrm{T}(s_0, {\bm s}_{j+1}){\bm C}^{-1}({\bm s}_{j+1},{\bm s}_{j+1}){\bm \epsilon}_{j+1} \\
  &= \mu(s_0) + 
  \begin{bmatrix}
  {\bm c}^\mathrm{T}(s_0, {\bm s}_j) & c(s_0, s_{j+1})
  \end{bmatrix}  \\
  & \quad \begin{bmatrix}
  {\bm C}({\bm s}_j, {\bm s}_j) & {\bm c}({\bm s}_j, s_{j+1})  \\
  {\bm c}^\mathrm{T}({\bm s}_j, s_{j+1}) & c(s_{j+1},s_{j+1})
  \end{bmatrix}^{-1}
  \begin{bmatrix}
  {\bm \epsilon}_j \\
  \epsilon_{j+1}
  \end{bmatrix}
  \end{split}
  \end{equation}
  where ${\bm \epsilon}_j=T_{gh}^{-1} \left( y(s_j)-\mu(s_j) \right)$ and  $\epsilon_{j+1}=T_{gh}^{-1} (y(s_{j+1})-\mu(s_{j+1}))$. \\
  ${\bm C}^{-1} ({\bm s}_{j+1}, {\bm s}_{j+1}){\bm \epsilon}_{j+1}$, which is a $(N +1) \times 1$ vectors that appears in Eq.(\ref{eq:mu}), has the following recursive expression (Eq.9 of \cite{gramacy2015local}):
  
  \begin{equation}\label{eq:K}
  \begin{split}
  &{\bm C}^{-1} ({\bm s}_{j+1}, {\bm s}_{j+1}){\bm \epsilon}_{j+1} = \\
  &\begin{pmatrix}
  {\bm C}^{-1} ({\bm s}_j, {\bm s}_j){\bm \epsilon}_j + {\bm g}_j(s_{j+1})[h_j(s_{j+1})\beta_j(s_{j+1})+\epsilon_{j+1}] \\
  h_j(s_{j+1})+\epsilon_{j+1}/\beta_j(s_{j+1})
  \end{pmatrix}
  \end{split}
  \end{equation}
  where ${\bm g}_j (s_{j+1}) = {\bm C}^{-1} ({\bm s}_j, {\bm s}_j){\bm c}( {\bm s}_j,s_{j+1} )/\beta(s_{j+1})$, $h_j (s_{j+1})= {\bm \epsilon}^\mathrm{T}_j {\bm g}_j (s_{j+1})$. $\beta_j(s_{j+1})$ is defined later. Using Eq.(\ref{eq:K}), $a(s_0)|{\bm y}({\bm s}_{j+1})$ can be updated sequentially as follows:
    
  \begin{equation}\label{eq:mu_sec}
  \begin{split}
  a(s_0)|{\bm y}({\bm s}_{j+1})= & a(s_0)|{\bm y}({\bm s}_j) + c(s_0, s_{j+1})[h_j(s_{j+1})\\
  &+\epsilon_{j+1}/\beta_j(s_{j+1})]
  \end{split}
  \end{equation}
  The sequential calculation does not involves the heavy inversion of ${\bm C}^{-1} ({\bm s}_{j+1}, {\bm s}_{j+1})$; therefore, it is computationally efficient.
  
  Regarding $b^2(s_0)$, Lemma 3 in \cite{xu2017tukey} showed that the variance under the TGH process is identical to the variance under GP. Therefore, in the case with a local TGH process, the conditional variance is given just like laGP as follows (Eq.5 of Gramacy and Apley, 2015):
    
 \begin{equation}\label{eq:V}
 b^2(s_0)|{\bm y}({\bm s}_{j+1})=\frac{\psi_j}{j-2}\beta_{j+1}(s_0)
\end{equation}
where $\psi_j={\bm \epsilon}^\mathrm{T}_j{\bm C}^{-1}({\bm s}_j, {\bm s}_j){\bm \epsilon}_j$ and 
\begin{equation}\label{eq:beta}
\beta_{j+1}(s_0)=c(s_0,s_0)-{\bm c}^\mathrm{T}(s_0, {\bm s}_{j+1}){\bm C}^{-1}({\bm s}_{j+1}, {\bm s}_{j+1})\\{\bm c}(s_0,{\bm s}_{j+1})
\end{equation}
The following recursive equation is derived based on \cite{gramacy2015local}:
    
    \begin{equation}\label{eq:v_sec}
  \begin{split}
  \beta_j(s_0)- & \beta_{j+1}(s_0) =\\
  & \quad \beta_j(s_{j+1}){\bm c}^\mathrm{T}(s_0,{\bm s}_{j+1}){\bf G}_j({\bm s}_{j+1}){\bm c}(s_0,{\bm s}_{j+1})\\
  & \quad +2c(s_0, s_{j+1}){\bm c}^\mathrm{T}(s_0, {\bm s}_{j+1}){\bm g}_j({\bm s}_{j+1})\\
  & \quad +\frac{c(s_0, s_{j+1})^2}{\beta_j(s_{j+1})}
  \end{split}
  \end{equation}
  \begin{equation}\label{eq:psi}
  \begin{split}
  \psi_{j+1}&=\psi_j + h_j(s_{j+1})^2\beta_j(s_{j+1})+2\epsilon_{j+1}h_j(s_{j+1}) \\
  & \quad +\epsilon^2_{j+1}/\beta_j(s_{j+1})
  \end{split}
  \end{equation}
  where ${\bm G}_j({\bm s}_{j+1})={\bm g}_j(s_{j+1}){\bm g}_j(s_{j+1})^\mathrm{T}$. Thus, $b^2(s_0)|{\bm y}({\bm s}_{j+1})$ is updated sequentially by substituting Eqs.(\ref{eq:v_sec}) and (\ref{eq:psi}) into Eq.(\ref{eq:V}).

 \subsubsection{Local sampling design}\label{sec:sample}
  
  Since the variance updating equations for the local TGH process are identical to the laGP, the following expression on the variance for laGP is available in our case:
  \begin{equation}
  b^2_{j+1}(s_0)|{\bm y}({\bm s}_{j+1}) = b^2_j(s_0)|{\bm y}({\bm s}_{j+1})-\frac{\psi_j}{j-2}R_{j+1}(s_0) \\
  \end{equation}  
  \begin{equation}
  \begin{split}
  &R_{j+1}(s_0) = \\
  &\quad \frac{\left[
  c(s_0,s_{j+1})-{\bm c}^\mathrm{T}(s_{j+1}, {\bm s}_j){\bm C}({\bm s}_j,{\bm s}_j)^{-1}{\bm c}(s_0, {\bm s}_j)
  \right]^2}{
    c(s_{j+1},s_{j+1})-{\bm c}^\mathrm{T}(s_{j+1}, {\bm s}_j){\bm C}({\bm s}_j,{\bm s}_j)^{-1}{\bm c}(s_{j+1}, {\bm s}_j)
  } \\
    \end{split}
    \end{equation}
    Then, the following local design algorithm, which is proposed by \cite{Sung2018exploit}), is available (reference: Algorithm 1 of \cite{Sung2018exploit}):
    
    \begin{enumerate}
    \item Let ${\bm s}_{k(-j)}$ denote the $k$ nearest neighbors to $s_0$ in the locations not currently in the sub-design, ${\bm s}_j$. Set $\delta_{j+1}$ as the maximum variance reduction from $N_{jk}(s_0)$. That is,
    \begin{equation}
    \delta_{j+1}=max_{u \in {\bm s}_{k(-j)}} R_u(s_0),
    \end{equation}
    \item Set \\ $z=\Phi^{-1} \left( \sqrt{\frac{\delta_{j+1}}{
    (1+\sqrt j \| {\bm C}({\bm s}_j, {\bm s}_j)^{-1}{\bm C}({\bm s}_j, s_0)\|_2)^2+
    j\delta_{j+1}/\lambda_{min}
    }} \right)$, \\ where $\lambda_{min}$ is the minimum eigenvalue of ${\bm C}({\bm s}_j, {\bm s}_j)$. Let
    \begin{equation}
    \begin{split}
    T({\bm s}_j) =& \{ u \in {\bm s}_{k(-j)} : \\
    & \quad \| u-v \|_2 \leq z \ {\rm for \  some } \  v \in \{s_0, {\bm s}_j \} \}
    \end{split}
    \end{equation}
    Then,
    \begin{equation}
    s_{j+1}= \argmax_{u \in T({\bm s}_j)} R_u(s_0),
    \end{equation}
    \item Set $j=j+1$ and repeat 2 and 3 until either the reduction in variance $R_u(s_0)$ falls below a prespecified threshold or the local design budget is met.
    \end{enumerate}
    
\subsubsection{Local TGH process estimation procedure}\label{sec:laGHest}
    The following procedure of \cite{gramacy2015local} may be applicable if only the laGP parameter estimation part is replaced with the laGH estimation:
    
    \begin{enumerate}
    \item Choose a sensible starting global lengthscale parameter ${\bm \theta_0}$ for all sites $s_0$.
    \item Calculate local design ${\bm s}_j$ for each $s_0$ based on sequential application of the procedure above independently.
    \item Furthermore, independently calculate the MLE lengthscale $\hat{\bm \theta}(s_0)|{\bm s}_j$, thereby explicitly obtaining a global nonstationary predictive surface.
    \item Set ${\bm \theta}(s_0) = \hat{\bm \theta}(s_0)$ possibly after spatially smoothing over all $s_0$ locations.
    \item Repeat steps 2--4 as desired. Then independently output each prediction $s_0$ based on ${\bm s}_j$ and possibly smoothed ${\bm \theta}(s_0)$.
    \end{enumerate}

Unlike the low rank approach, the sparse approach estimates local variations of skewness and kurtosis. Furthermore, the independent computation for each site enables parallelizing of local model estimations.

We deployed a Monte Carlo simulation experiment to examine the parameter estimation accuracy of the local TGH modeling approach. The result suggests that our approach properly estimates a positive $g$ value if the true process has a positive $g$. The same is true for negative $g$ values. Yet, the estimates have a relatively large variation because only a number of samples are used for the estimation . The local averaging, which we assume in the fourth step of the local TGH-RF estimation procedure, will be valuable to reduce the variation so that credible estimates can be obtained. 

Regarding the $h$ parameter, the estimate properly takes values near zero when the true process has a zero $h$ value, while the estimate tends to take larger values when the true process has a positive $h$. However, the estimates are smaller than the true value. Estimation of the $h$ parameter using small local samples would be an important next topic. See Supporting Material B for further detail.

\subsubsection{Application to the enumerated ground temperatures}\label{sec:laGHap}

This section applies the sparse TGH-RE approach to estimate the $\{a, b, g, h \}$ parameters at each site. To analyze the typical spatial patterns of these parameters, and also to stabilize these estimates, the sparse TGH approach is fitted by 5 day intervals from July 1, and the resulting parameter estimates are averaged over the days. For fast computation, the size of the local samples, which are optimized by the procedure explained in the previous section, is constrained to be equal to or less than 200. 

\begin{figure}
\begin{center}
\includegraphics[scale=0.35]{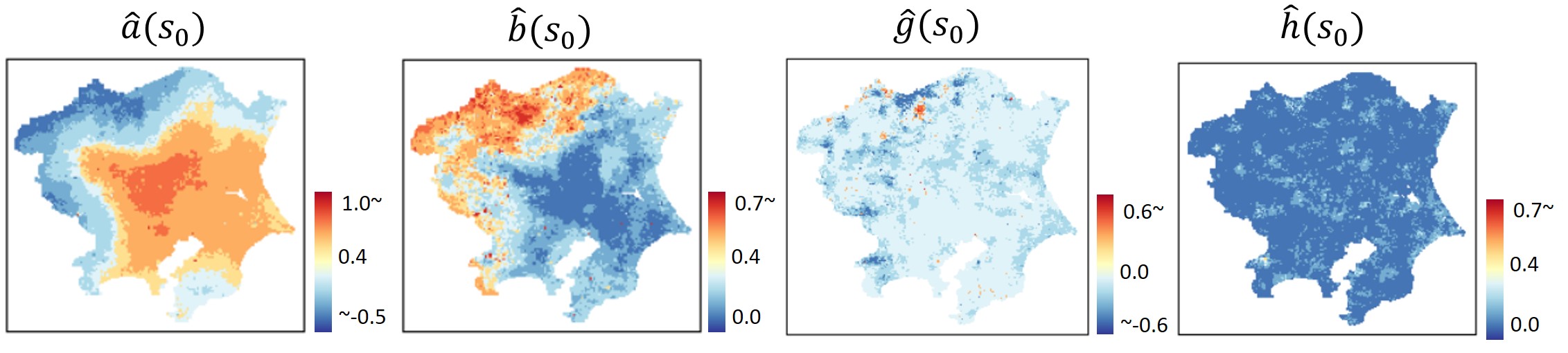}
\caption{Estimated parameters averaged across the 11 days.\label{fig:12}}
\end{center}
\end{figure}

\begin{figure}
\begin{center}
\includegraphics[scale=0.35]{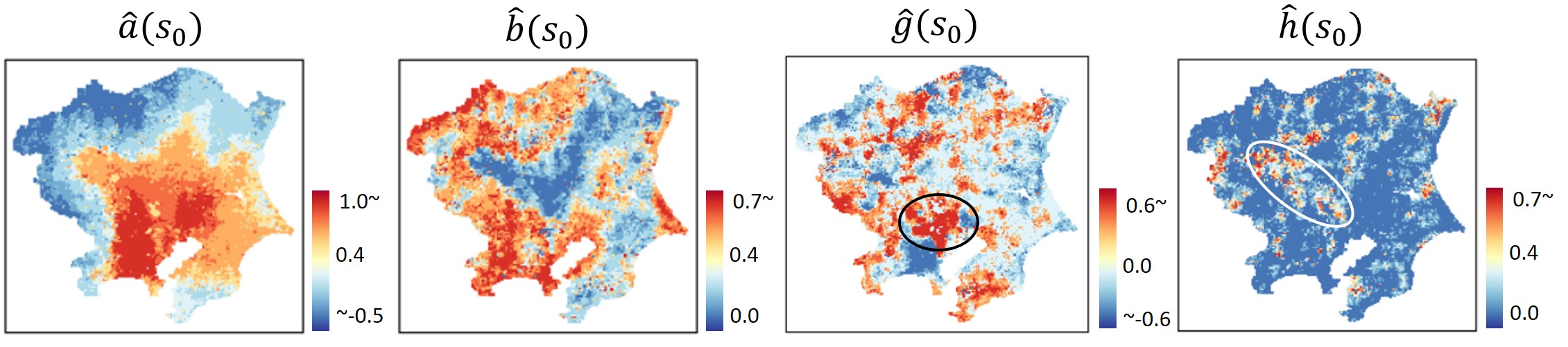}
\caption{Estimated parameters on July 19 when the mean of the maximum temperature is the highest. Black circuit shows the center of Tokyo while while shows an inland area, which is famous for heat. \label{fig:13}}
\end{center}
\end{figure}

Figure \ref{fig:12} summarizes the estimated parameters averaged across the 11 days that is the target days except for July 19 indicating the mean maximum temperature. The estimated $\hat{a}(s_0)$ parameters indicate that temperatures in urban areas are higher than those in mountain areas. The estimated $\hat{b}(s_0)$ values indicate large values in mountain areas. It is known that the up- and down-drafts generated from mountain make weather uncertain. The result confirms that our approach captures such a local uncertainty in temperatures. On the other hand, $\hat{g}(s_0)$ and $\hat{h}(s_0)$ values are nearly zero across the region. These values are almost zero across the 11 days. Skewness and Kurtosis are not so influential at least in typical summer days.

Figure \ref{fig:13} plots the estimated parameters in July 19 when the mean maximum temperatures was the highest across the target days. Interestingly, their spatial patterns are quite different from the average patterns shown in Figure \ref{fig:12}. The $\hat{a}(g_0)$ parameter shows similar spatial pattern in mountain areas but much higher values in the near center. This result verifies the presence of the urban heat island effect near the center. $\hat{b}(g_0)$ takes large values in coastal areas near the center. It suggests that sea breeze, humidity, and sea-related factors make temperature uncertain in hot days. The $\hat{g}(g_0)$ parameter indicates large positive values in the center of Tokyo, which are shown by a black circle. It is suggested that the extreme heat means not just temperature increase but also change of temperature distribution shape that determines risk (e.g., exceedance probability). $\hat{g}(g_0)$ values tend to take positive values in outer mountain areas too. Consideration of skewness is especially important in hot days. 

The $\hat{h}(s_0)$ parameter tends to indicate high values in a basin area shown in a white circle. This area is known as a hot area because the air warmed in the center flow into this basin. The large value indicating a fat tail might be because temperature significantly changes depending on hot air comes or not from the center. The result that a fat tail (or a large $\hat{h}(s_0)$ value) appears only in intensively heat day will also be important to evaluate probabilistic risk appropriately.

\section{Discussion}\label{sec:4}

The previous section analyzes the spatial variation in the skewness and kurtosis in ground temperatures. This section further explore distribution properties mainly focusing on temporal aspects while considering spatial heterogeneity of temperature distributions. 

\subsection{Outline}\label{sec:4_1}

Based on the result that skewness and kurtosis changes over space, we divided the study area into seven sub-regions\footnote{Although we first attempted to optimize the number of sub-regions with BIC minimization, the resulting number of cluster was too large, and difficult to interpret. Hence, we used seven clusters, which were interpretable (See Supporting Material B), and distributional properties were analyzed in each of the sub-regions.} The clustering was performed as follows:
\begin{description}
\item[(i)] The TGH distribution is fitted on $\{ \tilde{Y}_1(s_i), \ldots, \tilde{Y}_{61}(s_i)\}$, where 1-61 represents the index of the 61 days, to estimate the moment parameters $\{ a(s_i), b(s_i), g(s_i), h(s_i) \}$ in each grid $s_i$. We use the $l$-moment matching method of \cite{peters2016estimating}. They showed that their method estimates the moment parameters more robustly and accurately than classical moment matching, maximum likelihood, and quantile matching methods.
\item[(ii)] $k$-means clustering is applied to $\{a(s_i) , b(s_i), g(s_i), \\ h(s_i) \}$, and the 30,572 grids are grouped into seven sub-regions. We use a standard $k$-means method that defines the $k$-mean centers by the arithmetic means of sub-samples.
\end{description}
The next subsection explains the $l$-moment matching method. The subsequent subsection explains the estimation results of the $l$-moment parameters. After that we explain the result of spatial clustering based on the estimated moment parameters.

\subsection{l-moment matching}\label{sec:4_2}

$l$-moments \cite{hosking1989some} are defined by linear combinations of order statistics. The $m$-th $l$-moment for a sample of ordered observations $\{ Y_1 \leq \ldots \leq Y_j \leq \ldots \leq Y_n\} $ is defined as
\begin{align} \label{eq:poplmom0}
l_m = \frac{1}{m} \sum_{j=0}^{m-1} (-1)^i
\begin{pmatrix}
m-1 \\ i
\end{pmatrix}
E[Y_{m-i}]
\end{align}
$l$-moments exists if and only if the distribution has a mean; thus, $l$-moments can characterize a wider range of distributions than classical moments (\cite{peters2016estimating}). The first four $l$-moments have the following expressions:

\begin{align}
 l_1&=\int_0^1 F^{-1}_{y}(u)du,  \label{eq:poplmom1} &\\
 l_2&=\int_0^1 F^{-1}_{y}(u)(2u-1)du,  \label{eq:poplmom2} &\\
 l_3&=\int_0^1 F^{-1}_{y}(u)(6u^2-6u+1)du,  \label{eq:poplmom3} &\\
 l_4&=\int_0^1 F^{-1}_{y}(u)(20u^3-30u^2+12u-1)du,  \label{eq:poplmom4} &
\end{align}
where $F_y(u)$ is the distribution function for the random variable $y$ generated at quantile level $u$. Because $l_3$ and $l_4$ depend on the scale $l_2$, \cite{hosking1989some} suggested using the population $l$-skewness $\tau_3 =l_3/l_2$ and the population $l$-kurtosis $l_4/l_2$ to measure skewness and kurtosis. 

Unbiased estimators for the $l$-moments, which are also called sample $l$-moments, are given by
\begin{align}
 \hat{l_1} &= q_0,  \label{eq:slmom1} &\\
 \hat{l_2} &= 2 q_1 - q_0,  \label{eq:slmom2} &\\
 \hat{l_3} &= 6 q_2 - 6 q_1 + q_0,  \label{eq:slmom3} &\\
 \hat{l_4} &= 20 q_3 - 30 q_2 + 12 q_1 - q_0,  \label{eq:slmom4} &
\end{align}
where $q_m$ equals
\begin{equation} 
q_m =
\begin{cases}
  \frac{1}{n} \sum_{j=1}^n Y_j & if m=0 \\
  \frac{1}{n} \sum_{j=1}^n \frac{(j-1)(j-2) \cdots (j-m)}{(n-1)(n-2) \cdots (n-j)} Y_j & m>0 \\
\end{cases}
\end{equation}
\cite{peters2016estimating} proposed a $l$-moment matching method to estimate the moment parameters in the TGH model. In this section, which assumes a general setting, $\{a(s_i) , b(s_i), g(s_i) ,h(s_i) \}$ are simplified as $\{a , b, g ,h \}$. The proposed estimation steps are as follows:
\begin{itemize}
\item[(i)]  $g$ and $h$ are estimated by solving the following problem:
\begin{equation}
\{ \hat{g}, \hat{h} \} = \argmin_{ \{  \tau_3, \tau_4 \} } (\tau_3 - \hat{\tau_3})^2 + (\tau_4 - \hat{\tau_4})^2
\end{equation}
where $\hat{\tau_3} =\hat{l_3}/\hat{l_2}$ $\hat{\tau_4} =\hat{l_4}/\hat{l_2}$. 
\item[(ii)]  The estimates of $a$ and $b$ given their unbiased estimator are
\begin{align}
\hat{b} &= \hat{l}_2/l_2 \\
\hat{a} &= \hat{l}_1 - \hat{b} l_1.
\end{align}
\end{itemize}
For further details of the estimation, see \cite{peters2016estimating}. 

\begin{figure}
\begin{center}
\includegraphics[scale=0.35]{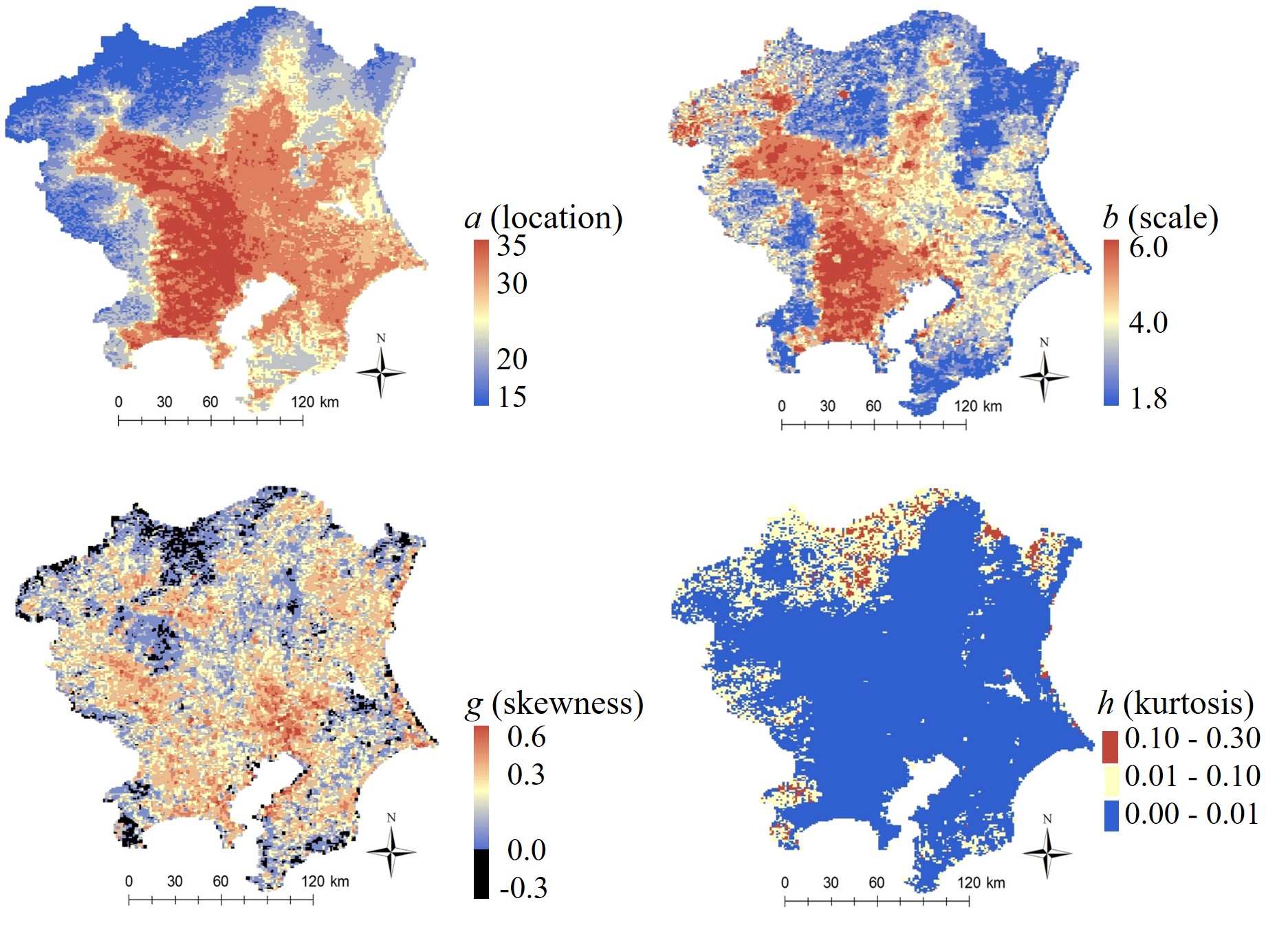}
\caption{Estimated $\{a(s_i), b(s_i), g(s_i), h(s_i)\}$ \label{fig:5}. These parameters are estimated using the emulated ground temperatures at 13:00 in the 61 target days. For comparison, the legend value range for $a(s_i)$ is equated with the range  for ground temperatures in Figure \ref{fig:gtemp}. \label{fig:abgh}}
\end{center}
\end{figure}

\subsubsection{Estimation result}\label{sec:4_3}

This section applies the $l$-moment matching to the emulated ground temperatures at 13:00 during the 61 target days $\{ \tilde{Y_1}(s_i), \ldots, \tilde{Y_{61}}(s_i)\}$ in each grid, and the moment parameters are estimated. Figure \ref{fig:A1} shows plots of the estimated $\{\hat{a}(s_i), \hat{b}(s_i), \hat{g}(s_i), \hat{h}(s_i)\}$ This figure demonstrates that these parameters change across regions. Both $\hat{a}(s_i)$ and $\hat{b}(s_i)$ inflate in highly urbanized areas, including the center and the west side. It is suggested that urbanization increases not only the mean temperatures but also the temperature variations, which can cause extreme heat. Unlike the ground temperatures plotted in Figure \ref{fig:A1}, $\hat{a}(s_i)$ takes the highest value in the west side of the center. It is reasonable because this area includes an inland area that is famous for experiencing high heat values. The estimated $\hat{g}(s_i)$ has positive values, which means it is right skewed, in most areas, which implies the possibility of having positive skewness in temperature, indicating higher temperatures are more probable. $\hat{h}(s_i)$ values are very small in many areas, though this is still significant as it indicates the probability of high temperatures in excess of what would be expected by normally distributed temperatures.

\subsubsection{Clustering result}\label{sec:4_4}
As discussed, to analyze the temperature behavior in each region, we divided the target area into seven clusters (see the Supporting Material A for a study on the effect of the number of clusters considered). The $k$-means clustering was applied to estimated, location, scale, skew and kurtosis related Tukey g-and-h parameters. $\{\hat{a}(s_i), \hat{b}(s_i), \hat{g}(s_i), \hat{h}(s_i)\}$. The result is plotted in the left panel of Figure \ref{fig:A2}. Note that the four parameters were standardized before the clustering as it eliminates scale dependency. For comparison, $k$-means method considering only mean temperature {$\hat{a}(s_i)$} was also applied, and plotted in the right panel of Figure \ref{fig:A2}. The right panel simply indicates the tendency that the central area is the hottest, and the temperature declines as the distance from the center increases. By contrast, when $\{\hat{a}(s_i), \hat{b}(s_i), \hat{g}(s_i), \hat{h}(s_i)\}$ are considered, the central area and the north are in the same cluster. The result is intuitively reasonable; these two areas are known as hazardous areas owing to the heatwave that is brought from central Tokyo by the south wind. It is suggested that consideration of not only mean, but also variance, skewness, and kurtosis reveals the hidden heatwave structure.

The goodness of the two clustering results are compared by Eq.(\ref{eq:cl}):

\begin{figure}
\begin{center}
\includegraphics[scale=0.35]{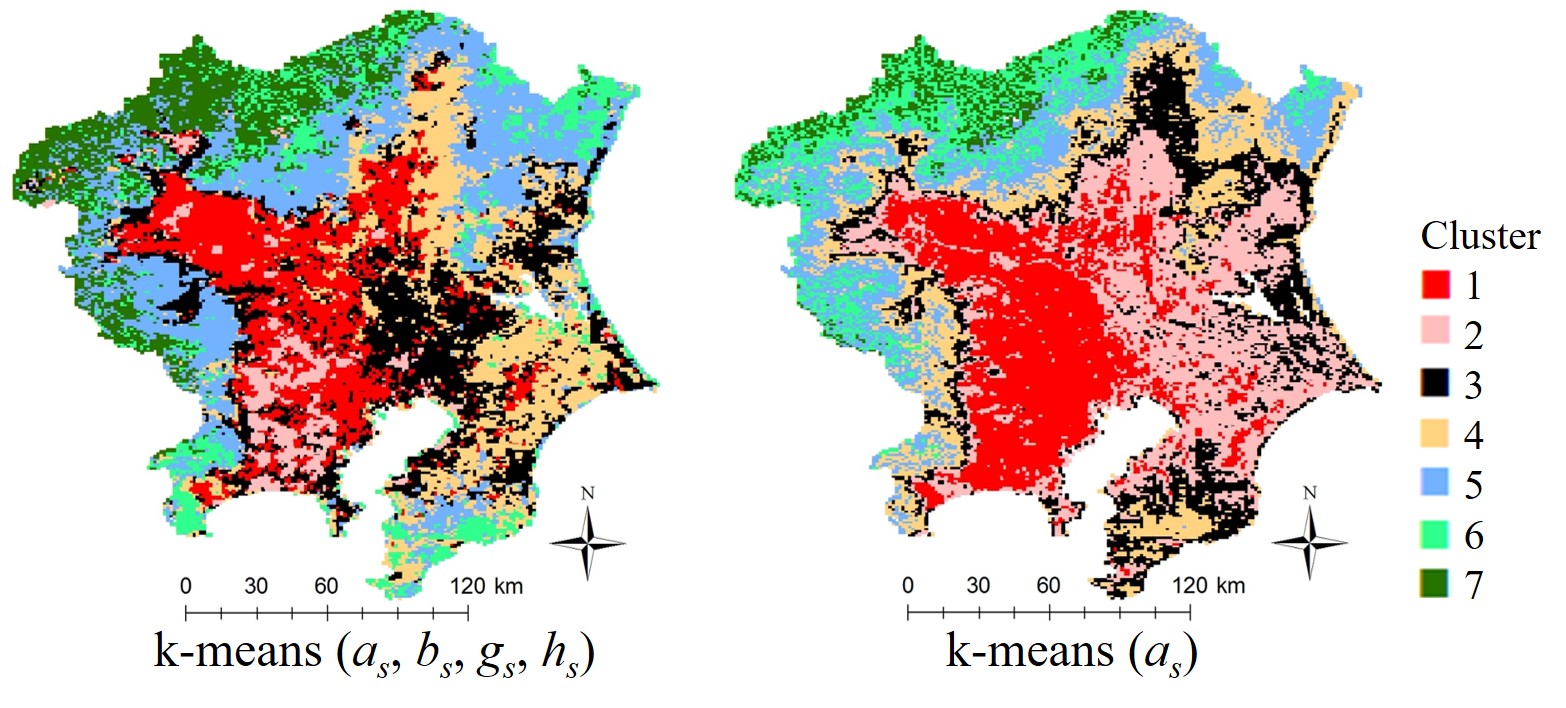}
\caption{Clustering result. Left: $\{a(s_i), b(s_i), g(s_i), h(s_i)\}$; right: {$a(s_i)$} \label{fig:6}}
\end{center}
\end{figure}

\begin{equation} \label{eq:cl}
D = \frac{1}{7^2} \sum^7_{C=1} \sum^7_{C^\mathrm{'}=1} \frac{ \sum_{s_i \in C} \sum_{s_i^\mathrm{'} \in C^\mathrm{'}} 
                  \frac{1} {|C||C^\mathrm{'}|}\|m(s_i) - m(s_i)^\mathrm{'}\|}
              { \sum_{s_i \in C} \sum_{s_i^\mathrm{'} \in C^\mathrm{'}} 
                  \frac{1} {|C|^2}\|m(s_i) - m(s_i)^\mathrm{'}\|}
\end{equation}
$m(s_i) \in \{\tilde{a}(s_i), \tilde{b}(s_i), \tilde{g}(s_i), \tilde{h}(s_i) \}$, where the four parameters equal $\{\hat{a}(s_i), \hat{b}(s_i), \hat{g}(s_i), \hat{h}(s_i)\}$ after the standardization, and $C$ denotes a cluster, and $|C|$ is the number of grids in the cluster. $D$ takes a large value if the seven clusters are well separated. $D$ becomes 1.05 for (a), while it becomes 0.86 for (b). It is found that consideration of not only $\hat{a}(s_i)$ but also $\{\hat{b}(s_i), \hat{g}(s_i), \hat{h}(s_i)\}$ is necessary to cluster ground temperatures accurately while considering skewness and kurtosis to determine heatwave risk (see Section \ref{sec:1}).

Figure \ref{fig:A3} displays the boxplots of parameters in each cluster. As plotted in Figure A2, the clusters are numbered in accordance with the distance from the center. This figure shows that the seven clusters are distinctive: overlaps of value ranges of the mean, variance, and skewness parameters are small across clusters. These three parameters tend to take higher values near the center and low value in the peripheral areas. It is revealed that the temperature distribution property greatly changes depending on the degree of urbanization.
\begin{figure}
\begin{center}
\includegraphics[scale=0.4]{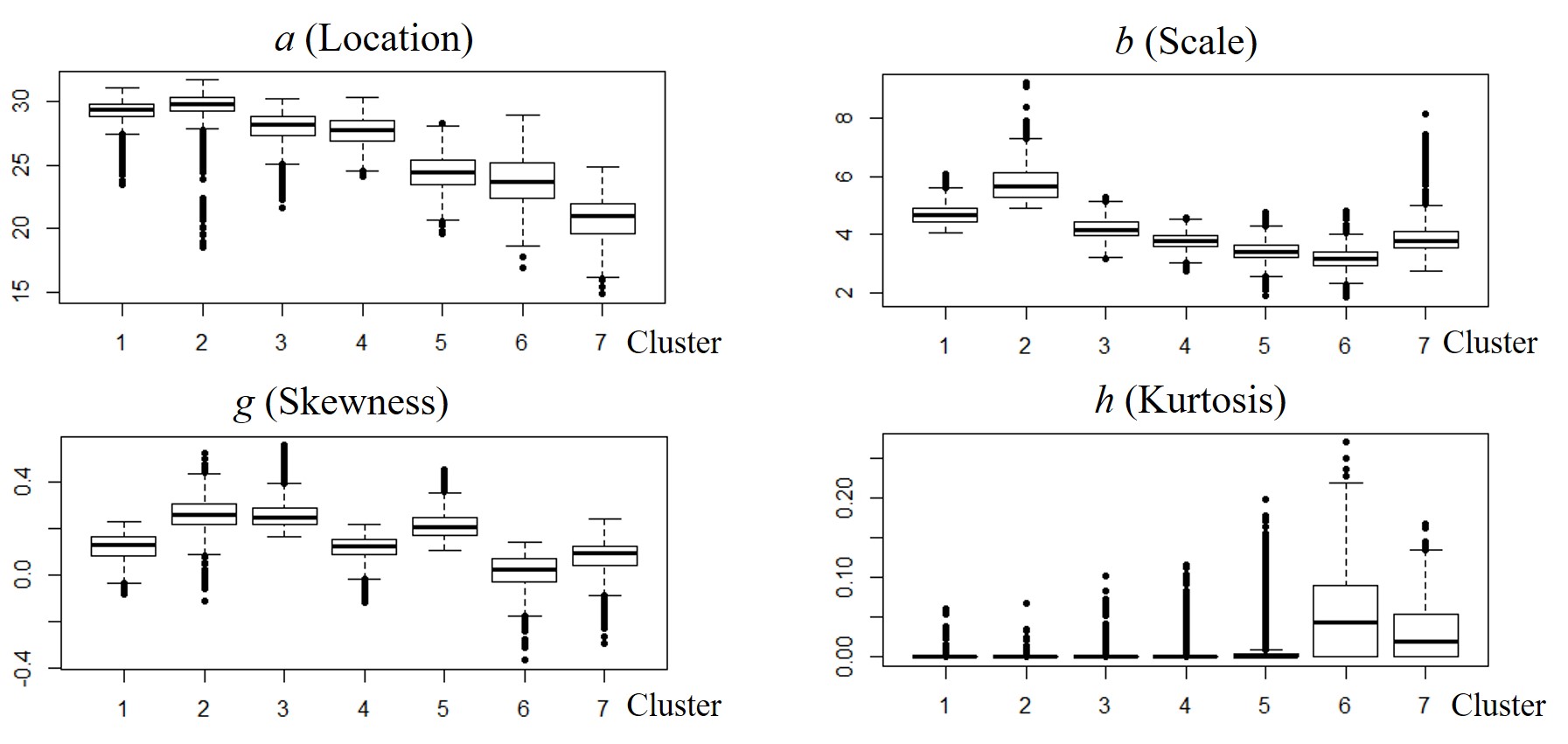}
\caption{Boxplots of the estimated parameters in each cluster. Cluster 1 is the nearest from the center, and cluster 7 is the farthest. \label{fig:7}}
\end{center}
\end{figure}

\section{Concluding remarks}\label{sec:6}

This study analyzes the spatial and temporal skew and kurtosis behavior of ground temperatures, which are emulated by combining ground and remotely sensed observations, using a $l$-moment matching approach for temporal analysis and low rank and sparse TGH-RF models for spatial analysis. These analysis results suggest the importance of considering skewness and kurtosis in heatwave modeling. Specifically, strong skewness values are estimated at the center of Tokyo and Kumagaya cities, which are known to experience a considerable amount of heat. The developed low rank and sparse TGH-RF models are found to be useful for revealing these tail structure from large spatial data.

Still, many remaining issues need to be addressed. First, we need to extend our developed low rank and sparse TGH models to dynamic spatio-temporal models. It is also important to quantify the heatwave "risk" that emerges when hazards (e.g., high temperature), exposure (e.g., many people are outside), and vulnerability (e.g., many of them are elders) are all set in the analysis. Fortunately, data relating these factors are increasingly available. For example, MODIS data quantifies heatwave hazards. Moreover, mobile GPS data can be used to determine how many people are exposed to the hazard. The national census dataset including the elderly ratio and the household income is valuable for estimating the ratio of people who are vulnerable to heat. District level heatwave risk estimation using a wide variety of micro spatial and temporal information would be an important step towards data-driven heat risk management, which is increasingly important as global warming advances.

\section{Acknowledgment}
This work was supported by JSPS KAKENHI Grant Number 17H01705.

\section{Supporting Material}
\subsection{Cluster analysis with different number of clusters}
This appendix performs a cluster analysis of the  parameters $\{ a(s_i), b(s_i), \\g(s_i), h(s_i)\}$ just like Section \ref{sec:4_4}, and shows that $n_C = 7$ is reasonable, where $n_C$ is the number of clusters.

Figure \label{fig:A1} plots the clustering result when $n_C$ is optimized by minimizing the Bayesian information criterion (BIC). $n_C=49$ becomes the optimal. However, the resulting clusters are extremely difficult to interpret. To obtain an interpretable clustering result, we restrict the number of clusters, $n_C$, at most 10.

Figure \label{fig:A2} plots AIC and BIC of the clusters when the number of clusters, $n_C$, equals 3, 4, $\ldots$ 10. AIC and BIC decrease as $n_C$ increase. However, the decrease becomes relatively slow after around $n_C = 7$. Figure \label{fig:A3} displays the clustering results when $n_C$ equals 3, 7, and 10. When $n_C$ = 3, the center area, and the suburban areas are merged in a cluster. However, central area is likely to have different heat behavior. Thus, the result with $n_C$ = 3 is too simple. Among the results when $n_C$ = 7 and 10, the former has a relatively clear spatial cluster pattern whereas the latter has a more mosaic-like pattern.

\renewcommand{\thefigure}{A1}
\begin{figure}
\begin{center}
\includegraphics[scale=0.25]{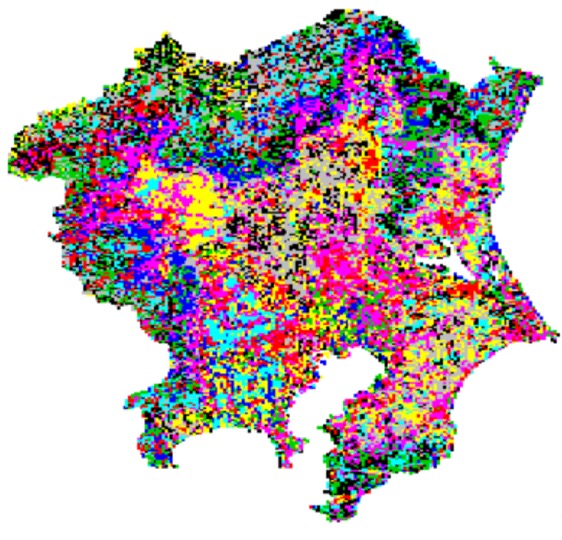}
\caption{The 49 clusters minimizing BIC}
\end{center}
\end{figure}

\renewcommand{\thefigure}{A2}
\begin{figure}
\begin{center}
\includegraphics[scale=0.5]{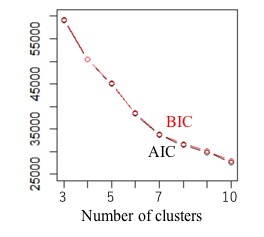}
\caption{Number of clusters (3 - 10) and AIC/ BIC}
\end{center}
\end{figure}

\renewcommand{\thefigure}{A3}
\begin{figure}
\begin{center}
\includegraphics[scale=0.35]{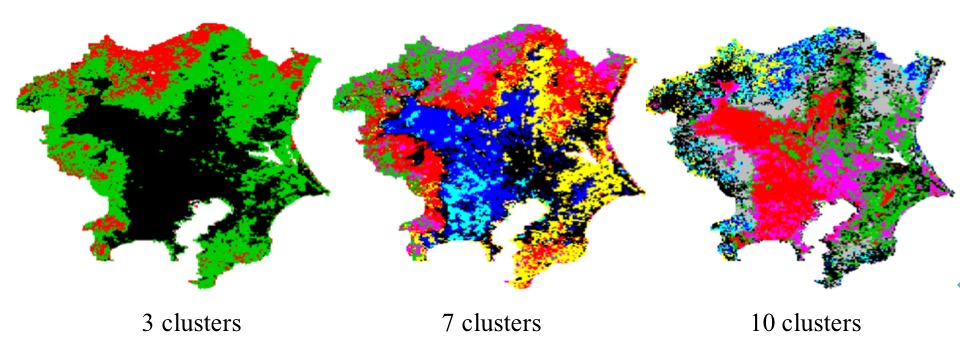}
\caption{Clustering results}
\end{center}
\end{figure}

Figure \ref{fig:A4} to \ref{fig:A7} are boxplots of the estimated a$(s_i)$, $b(s_i)$, $g(s_i)$, and $h(s_i)$ parameters in each of the clusters. The plots when $n_C = 3$ are quite different from the other two results. On the other hand, plots with $n_C  = 7$ and $n_C = 10$ have similar patterns.

Based on the aforementioned results, we use the 7 clusters whose spatial patterns are more clear than the 10 clusters (and the 49 clusters with minimum BIC), and boxplots of the parameters $\{ a(s_i), b(s_i), g(s_i), h(s_i)\}$ have similar pattern with the 10 clusters whose BIC is the minimum in cases with $n_C \in \{3, \ldots 10\}$.

\renewcommand{\thefigure}{A4}
\begin{figure}
\begin{center}
\includegraphics[scale=0.35]{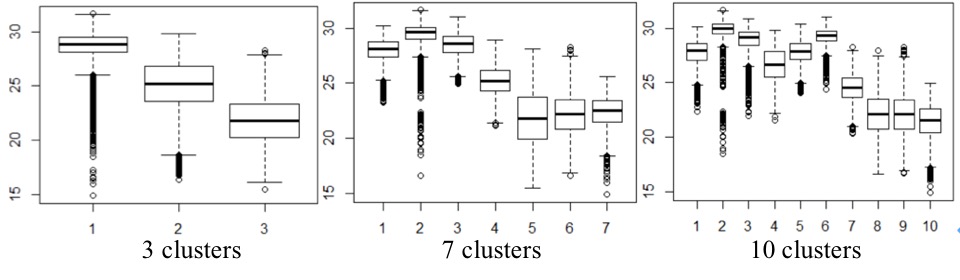}
\caption{Boxplots of the estimated $a(s_i)$ (mean) parameters in each cluster. The cluster 1 is the nearest from the center of Tokyo, the cluster 2 is the second nearest, and so on.}
\end{center}
\end{figure}

\renewcommand{\thefigure}{A5}
\begin{figure}
\begin{center}
\includegraphics[scale=0.35]{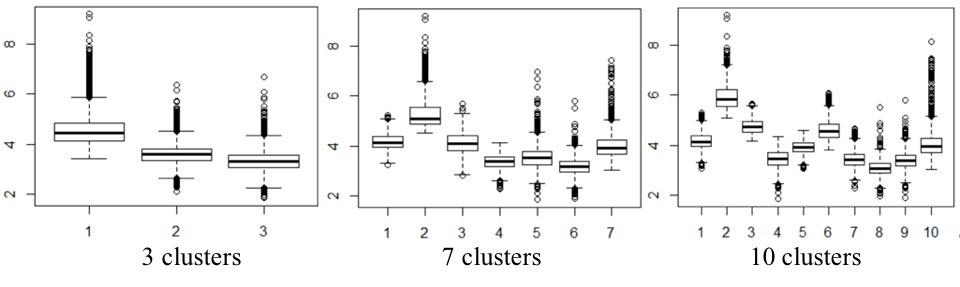}
\caption{Boxplots of the estimated $b(s_i)$ (scale) parameters in each cluster.}
\end{center}
\end{figure}

\renewcommand{\thefigure}{A6}
\begin{figure}
\begin{center}
\includegraphics[scale=0.35]{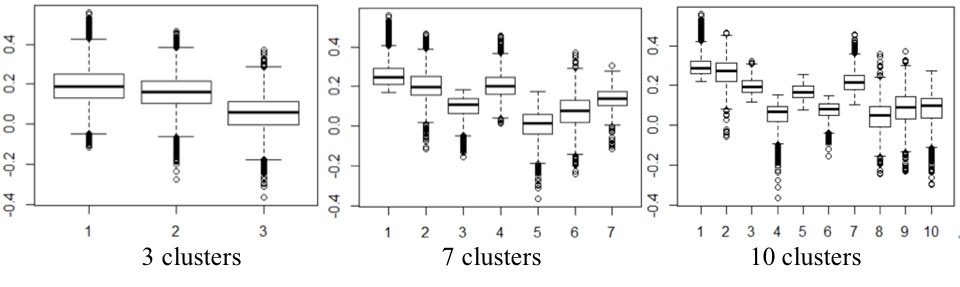}
\caption{Boxplots of the estimated $g(s_i)$ (skewness) parameters in each cluster.}
\end{center}
\end{figure}

\renewcommand{\thefigure}{A7}
\begin{figure}
\begin{center}
\includegraphics[scale=0.35]{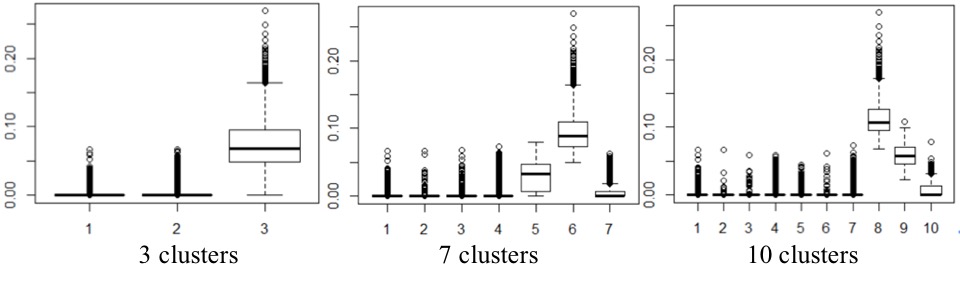}
\caption{Boxplots of the estimated $h(s_i)$ (kurtosis) parameters in each cluster.}
\end{center}
\end{figure}

\subsection{Simulation experiment on the low rank TGH-RF model}\label{sec:a2}
In this section, we compare our developed low rank TGH-RF model to the original TGH-RF model of \cite{xu2017tukey} in terms of computational time and estimation accuracy of {$a, b, g, h$}. 

Samples in July 1 in the cluster including the central area is used for the comparison. In the low rank approach, the number of eigen-pairs in our approximation is changed between 200 and 1200 by 200. All of our calculations are implemented in a Windows 10 64-bit system with 48 GB of memory, and coded using R (version 3.4.2).

Table \ref{tb:1} compares computational time. The computational time required for the low rank TGH-RF model estimation is much shorter time than the original model especially when $L$ is small. Interestingly, the low rank approach takes only 51 seconds even if $L = 1200$ while the full rank model, which implies $L =  n = 1500$, takes 3,494 seconds. The result demonstrates the computational efficiency of our approach that estimate the scale parameter $m$ instead of the range parameter $r$.

Figure \ref{fig:A8} summarizes estimated parameters. Parameters estimated from the low rank approach are reasonably similar when the number of eigen-pairs is equal or greater than 600.

In summary, the low rank approach accurately estimates the {$a, b, g, h$} parameters in a computationally efficient manner.

\renewcommand{\thetable}{A1}
\begin{table*}[!t]
 \caption{Comparison of computational time ($n = 1,500$)\label{tb:1}}
  \begin{center}
    \begin{tabular}{r r r} \\
\hline
Method & Number of eigenpairs: $L$ & CP time (second) \\
\hline
Xu and Genton (2016) & 1,500 & 3,493.92 \\
\hline
& 200 & 18.00 \\
& 400 & 23.71 \\
Ours & 600 & 29.02 \\
(average of 10 trials) & 800 & 40.50 \\
& 1000 & 49.60 \\
& 1200 & 50.97 \\
\hline
    \end{tabular}
  \end{center}
\end{table*}

\renewcommand{\thefigure}{A8}
\begin{figure}
\begin{center}
\includegraphics[scale=0.3]{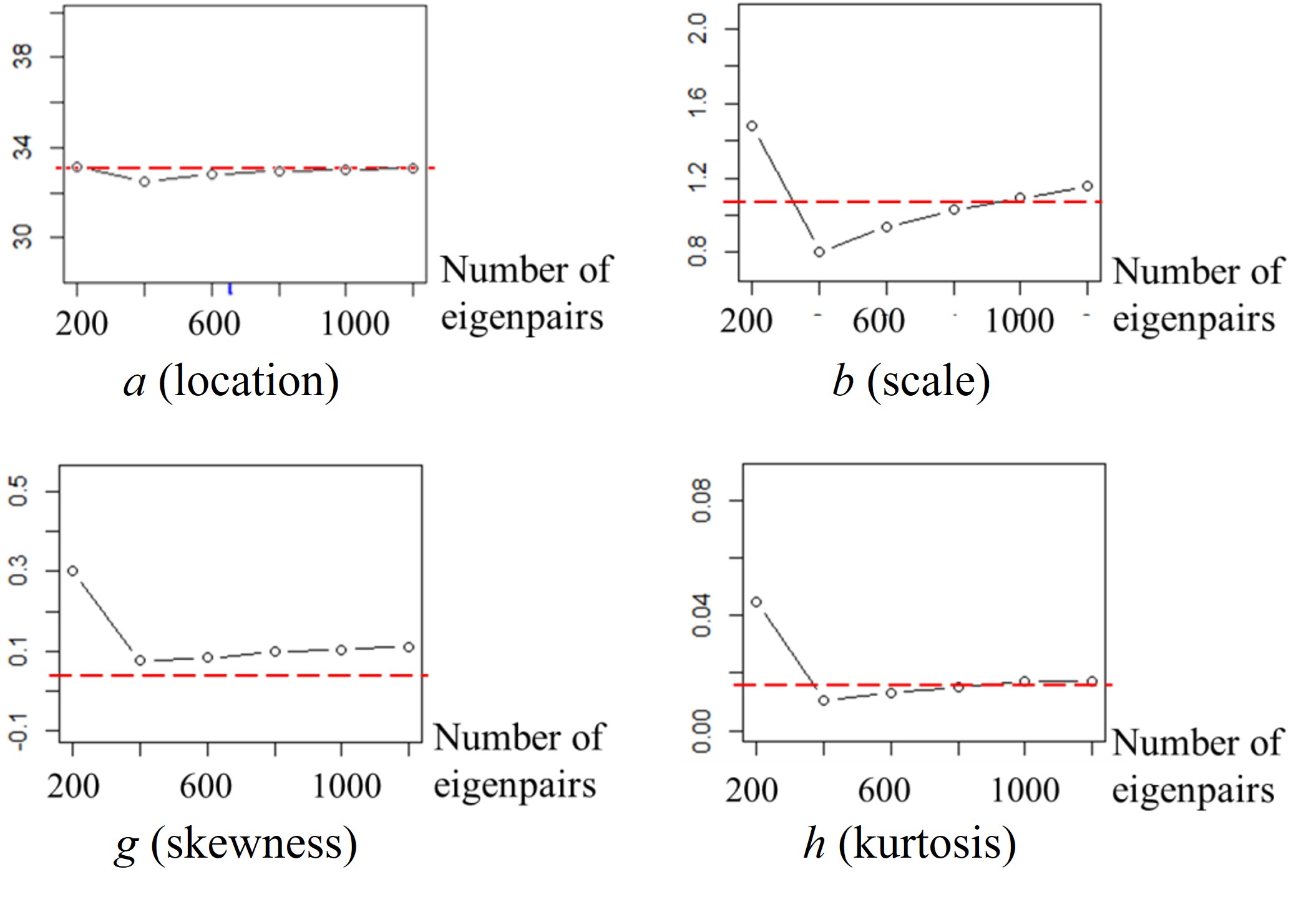}
\caption{Comparison of estimated parameters. Here parameters estimated in July 1 in the cluster, including the central area, are compared: Black: eigen-approximation; Red: exact. \label{fig:8}}
\end{center}
\end{figure}

\subsection{Simulation experiment on the sparse TGH-RF model}\label{sec:a3}
In this section, we examined the estimation accuracy of the sparse TGH-RF approach through a simulation experiment. TGH-RF is generated 100 times on 39 by 39 grids, and estimation accuracy of the $g$ and $h$ parameters at the center (20, 20) of the grid space is evaluated by fitting the sparse TGH-RF model. The center is denoted by $s_0$. Following Section \ref{sec:laGHap}, the sample size to estimate the model for $s_0$ is equal to or less than 200. The exponential kernel kernel $exp(-d(s_0,s_j)/r(s_0))$ is used to model spatial dependence where $r(s_0)$ is an unknown parameter. The true values for $a(s_0)$, $b(s_0)$, and $r(s_0)$ are given by 0, 1, and 1 respectively. The true values for the $g(s_0)$ and $h(s_0)$ parameters are specified as follows: $g(s_0) \in \{ -0.5, 0.0, 0.5\}$, $h(h_0) \in \{ 0.0, 0.25, 0.5 \}$; for each case, the sparse TGH-RF model is fitted 100 times.

Boxplots of the estimated $g(s_0)$ parameters are displayed in the top of Figure \ref{fig:Asim}. Our approach successfully detects a positive $g(s_0)$ value when the true value is positive. The same is true for negative $g(s_0)$. Although, the error tends to increase as $h(s_0)$ grows, the estimates are still unbiased. Our approach is found to detect skewness accurately. The bottom of Figure \ref{fig:Asim} shows boxplots for the estimated $h(s_0)$ parameters. The estimates are nearly unbiased across cases. The variance of the estimates are similar across cases. The accuracy of our approach is confirmed in terms of the kurtosis too.
 
\renewcommand{\thefigure}{A9}
\begin{figure}
\begin{center}
\includegraphics[scale=0.35]{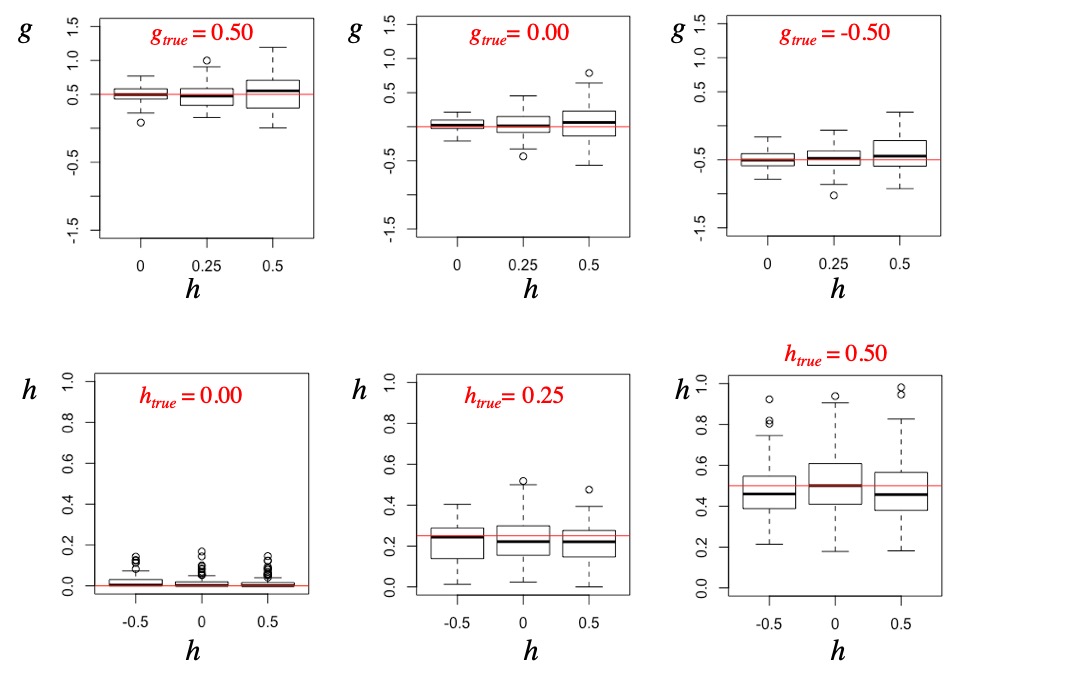}
\caption{Boxplots of the estimated $h(s_0)$ and $g(s_0)$ parameters. Red horizontal lines true parameter values. \label{fig:Asim}}
\end{center}
\end{figure}



\begin{thebibliography}{00}

\bibitem{oke1982energetic} Oke T. R., ``The energetic basis of the urban heat island,'' \emph{Quarterly Journal of the Royal Meteorological Society}, vo. 108, no. 455, pp. 1--24, Wiley Online Library, 1982.

\bibitem{semenza1996heat} Semenza J. C., Rubin C. H., Falter K. H, Selanikio J. D, Flanders W. D., Howe H. L., and Wilhelm J. L., ``Heat-related deaths during the July 1995 heat wave in Chicago,'' \emph{New England journal of medicine}, vo. 3, no. 2, Mass Medical Soc, 1996.

\bibitem{weng2004estimation} Weng Q., Lu D., and Schubring J., ``Estimation of land surface temperature--vegetation abundance relationship for urban heat island studies,'' \emph{Remote Sensing of Enviroment}, vo. 89, no. 4, pp. 467--483, Elsevier, 2004.

\bibitem{glaeser2010greenness} Glaeser E. L. and Kahn M. E., ``The greenness of cities: carbon dioxide emissions and urban development,'' \emph{Journal of Urban Economics}, vo. 67, no. 3, pp. 404--418, Elsevier, 2010.

\bibitem{mathews2010mobilizing} Mathews J. A., Kidney S., Mallon K., and Hughes M., ``Mobilizing private finance to drive an energy industrial revolution,'' \emph{Energy Policy}, vo. 38, no. 7, pp. 3263--3265, Elsevier, 2010.

\bibitem{buchner2011landscape} Buchner B. and Falconer A. and Herv{\'e}-Mignucci M. and Trabacchi C. and Brinkman M., ``The landscape of climate finance,'' \emph{Climate Policy Initiative, Venice}, vo. 27, 2011.

\bibitem{kidneygreening} Kidney S., Oliver P., Sonerud B., ``Growing a green bonds market in China,'' \emph{Climate Bonds}, 2015.

\bibitem{stewart2012local} Stewart I. D. and Oke T. R., ``Local climate zones for urban temperature studies,'' \emph{Bulletin of the American Meteorological Society}, vo. 93, no. 12, pp. 1879--1900, American Meteorological Society, 2012.

\bibitem{murakami2016participatory} Murakami D., Peters G. W., Yamagata Y., and Matsui T., ``Participatory sensing data tweets for micro-urban real-time resiliency monitoring and risk management,'' \emph{IEEE Access}, vo. 4, pp. 347--372, IEEE, 2016.
    
\bibitem{jorge1984some} Jorge M. and Boris I., ``Some properties of the Tukey g and h family of distributions,'' \emph{Communications in Statistics-Theory and Methods}, vo. 13, no. 3, pp. 353--369, Taylor \& Francic, 1984.

\bibitem{SaiGWPIN} Nagarajan S. G. and Peters G. W. and Nevat I., ``Spatial Field Reconstruction of Non-Gaussian Random Fields: The Tukey G-and-H Random Process,'' \emph{SSRN}, no. 3050592, 2018.
    
\bibitem{chen2017dynamic} Chen W. Y., Peters G. W., Gerlach R. H., and Sisson S. A., ``Dynamic Quantile Function Models,'' \emph{ArXiv}, no. 1707.02587, 2017.

\bibitem{xu2017tukey} Xu G. and Genton M. G., ``Tukey g-and-h random fields,'' \emph{Journal of the American Statistical Association}, vo. 112, no. 519, Taylor \& Francis, 2017.

\bibitem{batty2013big} Batty M., ``Big data, smart cities and city planning,'' \emph{Dialogues in Human Geography}, vo. 3, no. 3, SAGE Publications, 2013.

\bibitem{kitchin2014real} Kitchin R., ``The real-time city? Big data and smart urbanism,'' \emph{GeoJournal}, vo. 79, no. 1, pp. 1--14, Springer, 2014.
    
\bibitem{hancke2012role} Hancke G. P., Silva B. C., and Hancke G. P. Jr, ``The role of advanced sensing in smart cities,'' \emph{Sensors}, vo. 13, no. 1, pp. 393--425, Multidisciplinary Digital Publishing Institute, 2012.

\bibitem{cressie2008fixed} Cressie N. and Johannesson G., ``Fixed rank kriging for very large spatial data sets,'' \emph{Journal of the Royal Statistical Society: Series B (Statistical Methodology)}, vo. 70, no. 1, pp. 209--226, Wiley Online Library, 2008.
    
\bibitem{banerjee2008gaussian} Banerjee S., Gelfand A. E., Finley A. O., and Sang H., ``Gaussian predictive process models for large spatial data sets,'' \emph{Journal of the Royal Statistical Society: Series B (Statistical Methodology)}, vo. 70, no. 4, pp. 825--848, Wiley Online Library, 2008.
    
\bibitem{nychka2015multiresolution} Nychka D., Bandyopadhyay S., Hammerling D., Lindgren F., and Sain S., ``A multiresolution Gaussian process model for the analysis of large spatial datasets,'' \emph{Journal of Computational and Graphical Statistics}, vo. 24, no. 2, pp. 579--599, Taylor \& Francis, 2015.

\bibitem{furrer2006covariance} Furrer R., Genton M. G., Nychka D., ``Covariance tapering for interpolation of large spatial datasets,'' \emph{Journal of Computational and Graphical Statistics}, vo. 15, no. 3, pp. 502--523, Taylor \& Francis, 2006.
    
\bibitem{gramacy2015local} Gramacy R. B. and Apley D. W., ``Local Gaussian process approximation for large computer experiments,'' \emph{Journal of Computational and Graphical Statistics}, vo. 24, no. 2, pp. 561--578, Taylor \& Francis, 2015.
    
\bibitem{datta2016hierarchical} Datta A., Banerjee S., Finley A. O., and Gelfand A. E., ``Hierarchical nearest-neighbor Gaussian process models for large geostatistical datasets,'' \emph{Journal of the American Statistical Association}, vo. 111, no. 514, pp. 800--812, Taylor \& Francis, 2016.

\bibitem{kim2005analyzing} Kim H-M., Mallick B. K., and Holmes, C. C., ``Analyzing nonstationary spatial data using piecewise Gaussian processes,'' \emph{Journal of the American Statistical Association}, vo. 100, no. 470, pp. 653--668, Taylor \& Francis, 2005.

\bibitem{heaton2019case} Heaton M. J., Datta A., Finley A. O., Furrer R., Guinness J., Guhaniyogi Rajarshi., Gerber F., Gramacy R. B., Hammerling D., Katzfuss M., Lindgren F., Nychka D. W., Sun F., and Zammit-Mangion A., ``A case study competition among methods for analyzing large spatial data,'' \emph{Journal of Agricultural, Biological and Environmental Statistics}, vo. 24, no. 3, pp. 398--425.

\bibitem{cressie2010fixed} Cressie N., Shi T., and Kang E. L., ``Fixed rank filtering for spatio-temporal data,'' \emph{Journal of Computational and Graphical Statistics}, vo. 19, no. 3, pp. 724--745, Taylor \& Francis, 2010.

\bibitem{genton2007separable} Genton M. G., ``Separable approximations of space-time covariance matrices,'' \emph{Environmetrics}, vo. 18, no. 7, pp. 681--695, Wiley Online Libary, 2007.

\bibitem{cameletti2013spatio} Cameletti M., Lindgren F., and Simpson D., Rue H., ``Spatio-temporal modeling of particulate matter concentration through the SPDE approach,'' \emph{AStA Advances in Statistical Analysis}, vo. 97, no. 2, pp. 109--131, Springer, 2013.

\bibitem{smith1990max} Smith R. L., ``Max-stable processes and spatial extremes,'' \emph{Unpublished manuscript}, vo. 205, 1990.
    
\bibitem{kabluchko2009spectral} Zakhar K., ``Spectral representations of sum-and max-stable processes,'' \emph{Extremes}, vo. 12, no. 4, p. 401, Springer, 2009.
    
\bibitem{reich2012hierarchical} Reich B. J. and Shaby B. A., ``A hierarchical max-stable spatial model for extreme precipitation,'' \emph{The Annals of Applied Statistics}, vo. 6, no. 4, p. 1430, NIH Public Access, 2012.

\bibitem{buishand2008spatial} Buishand T. A., De Haan L., and Zhou C., ``On spatial extremes: with application to a rainfall problem,'' \emph{The Annals of Applied Statistics}, vo. 2, no. 2, pp. 624--642, Institute of Mathematical Statistics, 2008.

\bibitem{davison2012statistical} Davison A. C., Padoan S. A., and Ribatet M., ``Statistical modeling of spatial extremes,'' \emph{Statistical Science}, vo. 27, no. 2, pp. 161--186, Institute of Mathematical Statistics, 2012.

\bibitem{de2007extreme} De Haan L. and Ferreira A. F., \emph{Extreme Value Theory: An Introduction}, Springer Science \& Business Media, 2007.
    
\bibitem{yan2017non} Yan Y. and Genton M. G., ``Non-Gaussian autoregressive processes with Tukey g-and-h transformations,'' \emph{ArXiv}, no. 1711.07516, 2012.

\bibitem{xu2015efficient} Xu G. and Genton M. G., ``Efficient maximum approximated likelihood inference for Tukey's g-and-h distribution,'' \emph{Computational Statistics \& Data Analysis}, vo. 91, pp. 78--91, Elsevier, 2015.

\bibitem{gotway2002combining} Gotway C. A. and Young L. J., ``Combining incompatible spatial data,'' \emph{Journal of the American Statistical Association}, vo. 97, no. 458, pp. 632--648, Taylor \& Francis, 2002.

\bibitem{nevat2013random} Nevat I., Peters G. W., and Collings I. B., ``Random field reconstruction with quantization in wireless sensor networks,'' \emph{IEEE Transactions on Signal Processing}, vo. 61, no. 23, pp. 6020--6033, IEEE, 2013.

\bibitem{cressie2015statistics} Cressie N., \emph{Statistics for Spatial Data}, John Wiley \& Sons, 2015.
    
\bibitem{cressie1980robust} Cressie N. and Hawkins D. M., ``Robust estimation of the variogram: I,'' \emph{Journal of the International Association for Mathematical Geology}, vo. 12, no. 2, pp. 115--125, Springer, 1980.

\bibitem{kim2016projection} Kim D-W., Deo R. C., Chung J-H., and Lee J-S., ``Projection of heat wave mortality related to climate change in Korea,'' \emph{Natural Hazards}, vo. 80, no. 1, pp. 623--637, Springer, 2016.

\bibitem{lin2017climate} Lin C-Y., Chien Y-Y., Su C-J., Kueh M-T., and Lung S-C., ``Climate variability of heat wave and projection of warming scenario in Taiwan,'' \emph{Climatic Change}, vo. 145, no. 3-4, pp. 305--320, Springer, 2017.

\bibitem{okada2013proposal} Okada M. and Kusaka H., ``Proposal of a new equation to estimate globe temperature in an urban park environment,'' \emph{Journal of Agricultural Meteorology}, vo. 69, no. 1, pp. 23--32, The Society of Agricultural Meteorology of Japan, 2013.
   
\bibitem{hosking1989some} Hosking J. R. M., \emph{Some theoretical results concerning L-moments}, IBM Thomas J. Watson Research Division, 1989.

\bibitem{peters2016estimating} Peters G. W., Chen W. Y., and Gerlach R. H., ``Estimating quantile families of loss distributions for non-life insurance modelling via L-moments,'' \emph{Risks}, vo. 4, no. 2, p. 14, Multidisciplinary Digital Publishing Institute, 2016.
    
\bibitem{xu2016tukey} Xu G. and Genton M. G., ``Tukey max-stable processes for spatial extremes,'' \emph{Spatial Statistics}, vo. 18, pp. 431--443, Elsevier, 2016.
    
\bibitem{dray2006spatial} Dray S., Legendre P., and Peres-Neto P. R., ``Spatial modelling: a comprehensive framework for principal coordinate analysis of neighbour matrices (PCNM),'' \emph{Ecological Modelling}, vo. 196, no. 3-4, pp. 483--493, Elsevier, 2006.
        
\bibitem{schabenberger2017statistical} Schabenberger O. and Gotway C. A., \emph{Statistical Methods for Spatial Data Analysis}, CRC Press, 2017.

\bibitem{williams2001using} Williams C. K. I. and Seeger M. ``Using the Nystr{\"o}m method to speed up kernel machines,''  \emph{Advances in neural information processing systems}, 682--688, 2001.
    
\bibitem{smola2001sparse} Smola A. J. and Bartlett P. L., ``Sparse greedy Gaussian process regression,''  \emph{Advances in neural information processing systems}, 619--625, 2001.
    
\bibitem{peters2017statistical} Peters G. W., ``Statistical Machine Learning and Data Analytic Methods for Risk and Insurance,''  \emph{SSRN}, no. 3050592, 2017.

\bibitem{stein2014limitations} Stein M. L., ``Limitations on low rank approximations for covariance matrices of spatial data,'' \emph{Spatial Statistics}, vo. 8, pp. 1-19, Elsevier, 2014.

\bibitem{Sung2018exploit} Sung C-L., Gramacy R. B., and Haaland B., ``Exploiting Variance Reduction Potential in Local Gaussian Process Search,'' \emph{Statistica Sinica}, vo. 28, pp. 577--600, 2018.
        
\bibitem{gramacy2016speeding} Gramacy R. B. and Haaland B., ``Speeding up neighborhood search in local Gaussian process prediction,'' \emph{Technometrics}, vo. 58, no. 3, pp. 294--303, 2016.

\bibitem{nicholson2017bigvar} Nicholson W., Matteson D., and Bien J., ``BigVAR: Tools for Modeling Sparse High-Dimensional Multivariate Time Series,'' \emph{ArXiv}, no. 1702.07094, 2017.

\bibitem{peters2006bayesian} Peters G. W. and Sisson S., ``Bayesian Inference, Monte Carlo Sampling and Operational Risk,'' \emph{Journal of Operational Risk}, vo. 1, no. 3, 2006.

\bibitem{cruz2014fundamental} Cruz M. G., Peters G. W., and Shevchenko P. V., \emph{Fundamental Aspects of Operational Risk and Insurance Analytics: A Handbook of Operational Risk},'' John Wiley \& Sons, 2014.

\bibitem{nagarajan2018spatial} Nagarajan S. G. and Peters G. W., Nevat I., ``Spatial Field Reconstruction of Non-Gaussian Random Fields: The Tukey G-and-H Random Process,'' \emph{SSRN}, no. 3159687, 2018.

\end{thebibliography}
\end{document}